\documentclass[
aps,
twocolumn,
superscriptaddress,
showpacs,
english,
nofootinbib,
floatfix]
{revtex4-2}

\usepackage{mymacros}
\usepackage{graphicx}
\usepackage{placeins}
\usepackage{float}

\begin{document}

\title{Implications of the multi-minima character of molecular crystal phases onto the free energy}

\author{Marco Krummenacher} \affiliation{\UniBasel}
\author{Martin Sommer-Jörgensen} \affiliation{\UniBasel}
\author{Moritz Gubler} \affiliation{\UniBasel}
\author{Jonas A. Finkler} \affiliation{\UniBasel} \affiliation{\UniAalborg}
\author{Ehsan Rahmatizad Khajehpasha} \affiliation{\UniBasel}
\author{Giuseppe Fisicaro} \affiliation{\UniSicilia}
\author{Stefan Goedecker} \affiliation{\UniBasel}

\begin{abstract}
In recent years, significant advancements in computational methods have dramatically enhanced the precision in determining the energetic ranking of different phases of molecular crystals.
The developments mainly focused on providing accurate dispersion corrected exchange correlation functionals and methods for describing the vibrational entropy contributions to the free energy at finite temperatures. 
Several molecular crystals phases were recently found to have  of multi-minima character.
For our investigations we highlight the multi-minima character in the example of the molecular crystal consisting of N-(4’-Methylbenzylidene)-4-methylalanine.
We explore its potential energy landscape on the full DFT level or with a machine learned potential that was fitted to DFT data.
We calculate not only many local minima but also exact barriers along transformation pathways to demonstrate  the multi-minima character of our system.  
Furthermore, we present a framework, based on the quantum superposition method, that includes both configurational and vibrational entropy. 
As an example, we show for our system that the transition temperature between two of its phases is afflicted by an error of about 200 K if the multi-minima character is not taken into account. 
This indicates that it is absolutely essential to consider configurational entropy to obtain reliable finite temperature free energy rankings for complex molecular crystals.

\end{abstract}

\maketitle

\begin{figure*}
    \centering
    \includegraphics[width=18cm]{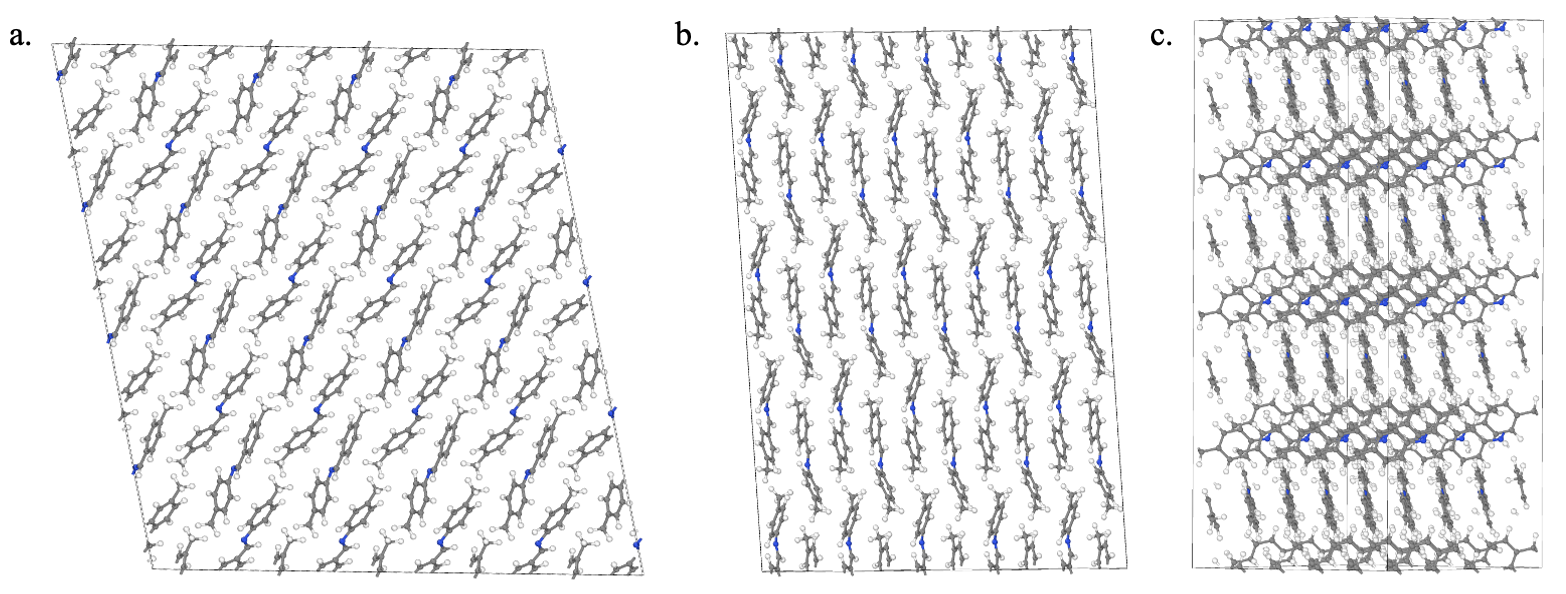}
    \caption{The three polymorphs of N-(4'-Methylkbenzylidene)-4-methylalanine: a. Form I, b. Form II and c. Form III}
    \label{fig:mc-forms}
\end{figure*}

Many organic substances occur in the form of molecular crystals (MCs), where discrete molecules are arranged in a periodic lattice structure. 
For most substances, several energetically similar crystallographic arrangements of these molecules exist. 
They are called polymorphs and in experiment they often differ only by 20 to 40 meV per atom~\cite{cruz2015facts}.
Understanding and predicting these different polymorphic forms is crucial since many physical properties, such as color, electronic conductivity, or solubility, are heavily dependent on the specific crystallographic arrangement of the molecules.\\
Given the uncertainty of having identified all pertinent solid forms experimentally, crystal structure prediction (CSP) is a valuable supplement to enhance the screening of different polymorphs and supplementing experiments~\cite{zhu2022analogy, cui2019mining, neumann2015combined, shtukenberg2017powder, braun2011racemic}.
The standard CSP approaches aim to locate all low energy local minima on the potential energy surface~\cite{day2011current, nyman2018crystal, thakur2015crystal,mortazavi2019computational, habgood2011form, stevenson2019solid}.
Because of the large number of required energy and force evaluations fast force fields are mainly used in this part.
The number of meta-stable configurations found in this way is typically much larger than the number of experimentally found polymorphs. 
This observation goes under the name of over-predicting. 
To get reliable structures together with their energetic ranking, the structures obtained from the force field are further refined with more accurate, but computationally demanding dispersion corrected DFT calculations~\cite{case2016convergence, neumann2008major, kronik2014understanding}. 
Many dispersion corrected schemes have been benchmarked, but it is still under debate which scheme gives the most accurate energetic ranking~\cite{hunnisett2024seventh}.
In standard CSP approaches, it is generally assumed that the structures lowest in the refined energies correspond to the experimentally observable polymorphs. 
At finite temperature, however, it is the free energy instead of the energy that determines the energetic ordering.
Vibrational entropy contributions are typically accounted for in calculations of the free energy, but configurational entropy has not yet been identified as an important factor 
for the finite temperature ranking of molecular crystals.
By running molecular dynamics simulations Dybeck et al.\ found that the trajectory is visiting the catchment basins of many local minima~\cite{dybeck2019exploring}.
Hence, a phase of a molecular crystal does not correspond to a single minimum on the PES but an ensemble of minima. 
The height of the barriers between the different minima belonging to one phase was then explored with the threshold algorithm which gives an upper limit to the barrier height~\cite{schon1996studying}.
Applying this algorithm to benzene, acrylic acid and resorcinol, Butler et al. found barriers in the range of meVs per atom~\cite{butler2023reducing}, confirming the MD observation that barriers can be overcome rapidly at room temperature. 
The same approach was later on also applied to more complex molecular crystals~\cite{yang2022global}. 
Other approaches to achieve such a clustering of the local minima were also proposed. 
Some of these methods were analyzing the packing similarity~\cite{montis2021combining}, the crystallization kinetics~\cite{beyer2001prediction, montis2020transforming} or are based on a series of MD runs followed by enhanced sampling simulations to group the minima of the PES into free energy basins~\cite{francia2020systematic, schneider2016exploring, sugden2022rationalising, song2020generating, francia2021reducing}. 
In this way the number or observable phases becomes much smaller than the number of local minima and the over-prediction conundrum is resolved.
The fact that the system can visit several catchment basins increases of course its entropy and lowers therefore its free energy. This can lead to a reversal in the free energy ordering of the phases where the phase higher in potential energy ends up having a lower free energy and thus is the preferred phase at finite temperature.
Similar effects exist also in standard inorganic materials~\cite{krummenacher2024entropic}, where configurational entropy can change the temperature of a solid-solid phase transition by about 200~K.\\
In the study presented here, we are quantifying with the help of the quantum superposition method~\cite{Walesbook} the influence of the multi-minima character of the molecular crystal consisting of N-(4’-Methylbenzylidene)-4-methylalanine on the transition temperature between its three experimentally observed phases.
This molecule was chosen due to its moderate molecular size, offering a balance between computational feasibility and structural complexity with its two connected carbon rings.
In addition, this molecular crystal belongs to a class of crystals which have shown the phenomenon of disappearing polymorphism where a seemingly stable polymorph suddenly transformed into another polymorph rendering the original polymorph extremely challenging or even impossible to reproduce.
Form I~\cite{bar1982molecular}, see Fig.~\ref{fig:mc-forms} a., was first described in 1973, but after it was stored for about eight months, the crystal samples did not diffract well.
Over three years recrystallization experiments were conducted, resulting in the discovery of Form II~\cite{bar1977molecular} (Fig.~\ref{fig:mc-forms}i b.) and 
Form III~\cite{bernstein1976molecular}   (Fig.~\ref{fig:mc-forms} c).
Nevertheless, the production of Form I was not successful. 
Recrystallization of Form I was achieved only after relocating to a laboratory with new equipment.

In our numerical investigation of this molecular crystal, we first performed Minima Hopping structure predictions~\cite{goedecker:2004,amsler, krummenacher2024performing,gubler2023efficient} for our molecular crystal. 
Given that classical force fields can lead to spurious minima on the potential energy surface~\cite{PhysRevB.81.214107}, we conducted these structure predictions at the DFT level using the Sirius DFT code~\cite{sirius} with an LDA functional that includes continuous first and second derivatives~\cite{goedecker1996separable}, thereby ensuring a very smooth potential energy surface. 
The LDA functional takes dispersion into account to some extent~\cite{price2021requirements} and it has proven to give the correct energetic ordering for many molecular crystals~\cite{beran2016modeling}. 
The minima were then re-relaxed using dispersion corrected DFT namely the D4 dispersion correction~\cite{caldeweyher2019generally, caldeweyher2020extension} combined with a PBE~\cite{perdew1996generalized} functional.
In order to allow the molecular crystal to adopt different space groups the Minima Hopping runs as well as all the other simulations were performed with variable cell shape. 
In this way 304 minima of all three known forms of N-(4’-Methylbenzylidene)-4-methylalanine were sampled.
DFT data extracted from our Minima Hopping simulations, were then also used to fit a NequIP~\cite{batzner20223} machine learning potential which was then later used as a calculator in the atomic simulation environment ASE~\cite{larsen2017atomic,krummenacher2024performing} to calculate vibrational frequencies  and to perform longer MD. These calculations would be too expensive on the DFT level.

\begin{figure}
    \centering
    \includegraphics[width=8.6cm]{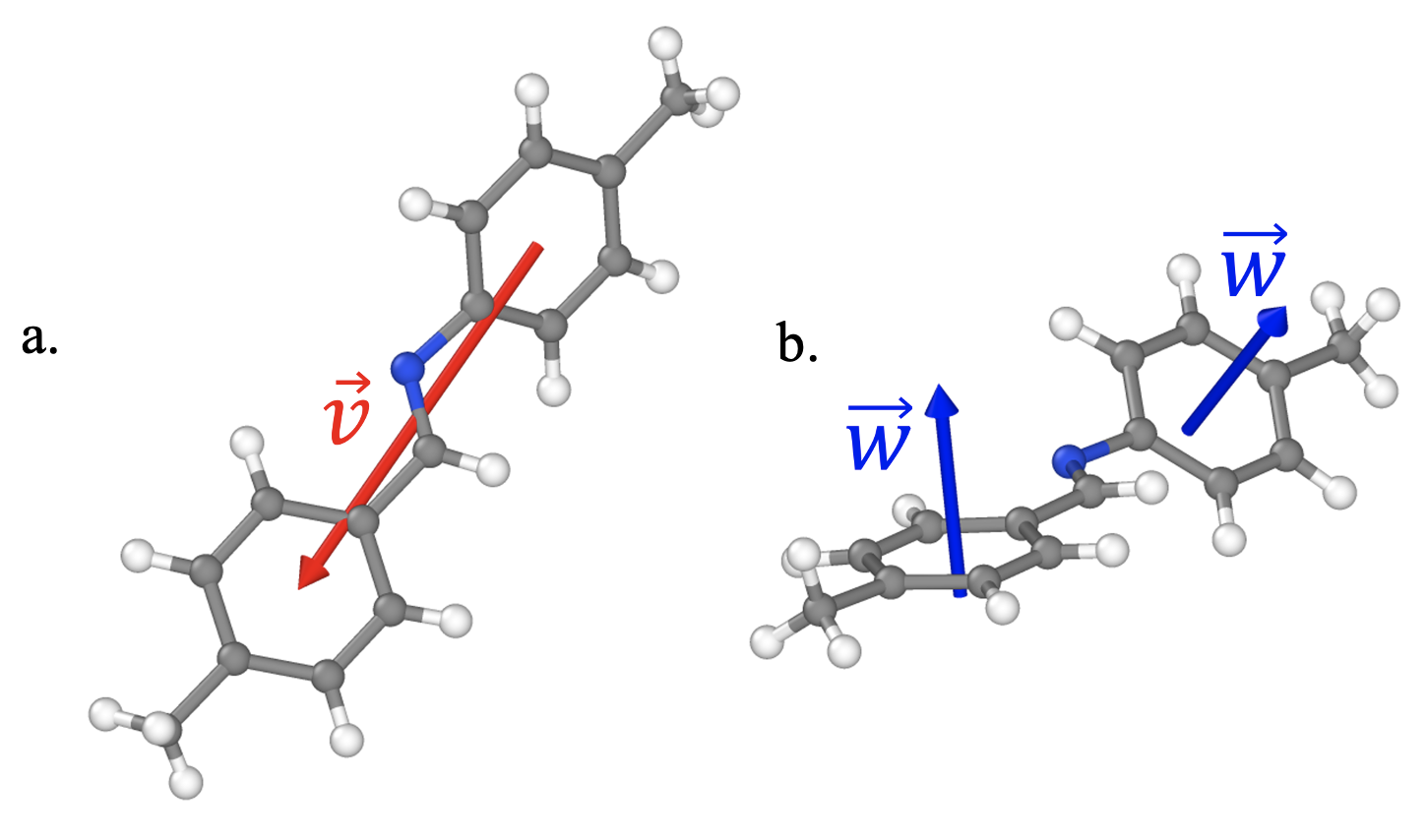}
    \caption{Vectors spanned to distinguish between the different polymorphs: a. To distinguish between Form III and Form II/I a vector $\Vec{v}$ between the carbon rings is defined as shown, b. Normal vectors $\Vec{w}$ through the surface spanned by the carbon rings were calculated to distinguish between Form II and Form I.}
    \label{vectors}
\end{figure}

Two collective variables were identified that allow us to classify the structures into the three experimentally found forms. 
To distinguish between Form III and Form I/II a vector $\Vec{w}$ was spanned between the center of mass of the two carbon rings of all the molecules in the unit cell. 
Then the angles between these vectors, each corresponding to a molecule, were calculated

\begin{equation}
    \alpha_{k}^{q} = \arccos(\frac{\Vec{v}_{i} \cdot \Vec{v}_{j}}{|\Vec{v}_{i}| |\Vec{v}_{j}|}),
\end{equation}
where $i$ and $j$ are the indices for the molecules of configuration $q$, $i \neq j$, and $\Vec{v}$ is the vector shown in Fig~\ref{vectors} a. 
Index $k$ counts all pairs $i,j$ and therefore goes from $1, \dots, \frac{N(N-1)}{2}$, where $N$ is the number of molecules in the unit cell of one configuration.
The resulting vector $\alpha_{k}^{q}$ 
consisting of all angles obtained within one structure is sorted to be permutationally invariant. 
To calculate a distance, this fingerprint vector is compared to the lowest energy Form III structure,

\begin{equation}
    d^{\alpha}(p,q) = \sum_{k} |\alpha_{k}^{p} - \alpha_{k}^{q}|^{2},
\end{equation}
where $\alpha^{p}_{k}$ are the sorted angles of the lowest energy reference structure of Form III.\\
For the categorization of the minima of Form I and Form II, we made use of the normal vectors of the carbon rings, see Fig.~\ref{vectors} b. 
For each carbon ring in the system, first, a plane was spanned and the normal vector of this plane was calculated by minimizing its distance to all carbon atoms of the carbon ring. 
Next, the angles of the normal vectors were calculated between all the carbon rings, except if it belonged to the same molecule

\begin{equation}
    \beta_{k}^{q} = \arccos(\frac{\Vec{w}_{i} \cdot \Vec{w}_{j}}
    {|\Vec{w}_{i}| |\Vec{w}_{j}|}),
\end{equation}

were $i$ and $j$ are the indices for the molecules of configuration $q$, $i \neq j$, and $i, j$ do not belong to the same molecule.
Index $k$ goes again from $1, \dots, N^{2}$.
The vector consisting of the angles was then compared to the reference structures of Form I and Form II,
\begin{equation}
    d^{\beta}(p,q) = \sum_{k} |\beta_{k}^{p} - \beta_{k}^{q}|^{2},
\end{equation}
where $\beta^{p}_{k}$ are the angles obtained from the reference structures of Form II or I, respectively.

\begin{figure}
    \centering
    \includegraphics[width=8.6cm]{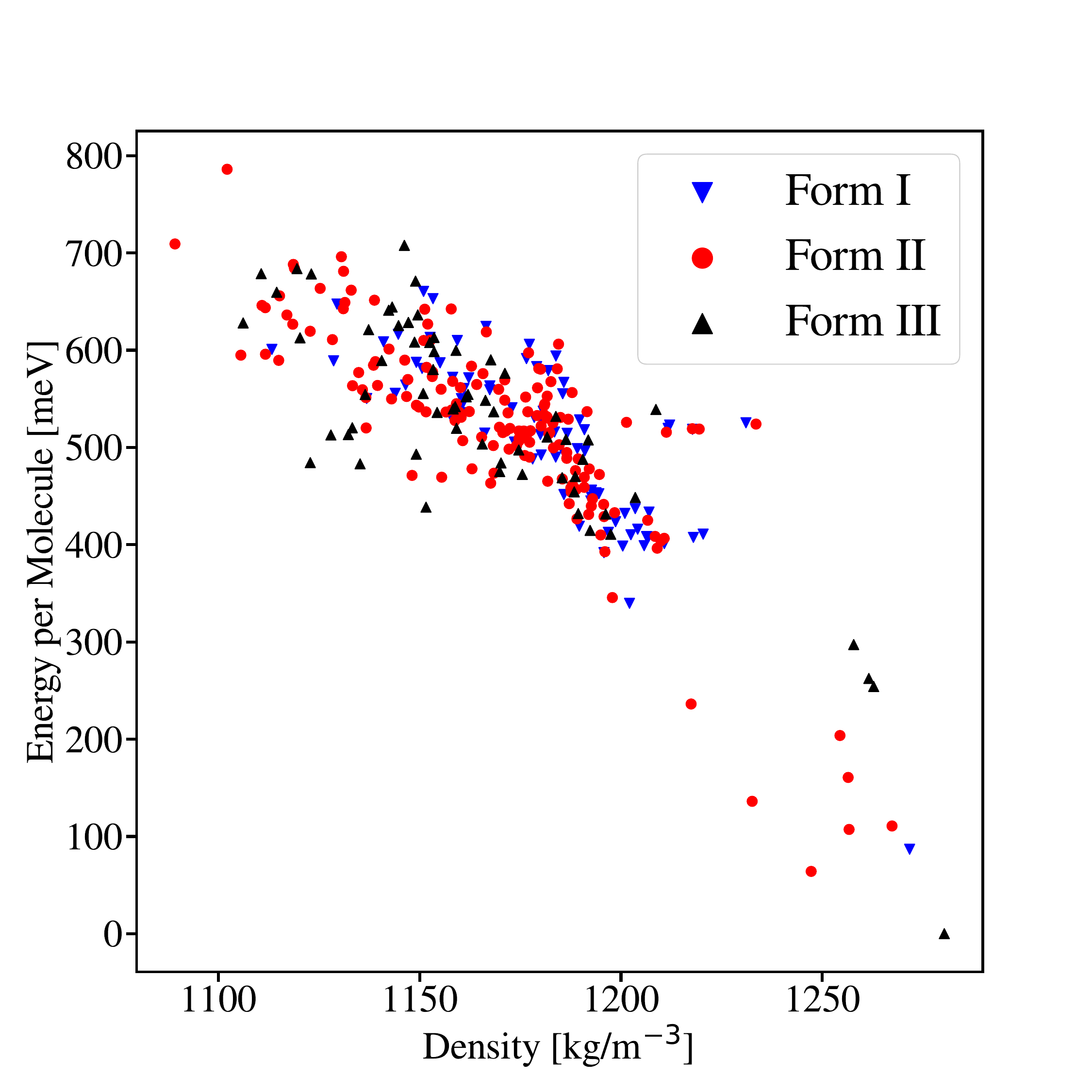}
    \caption{Energy-density diagram of the 304 structures found with minima hopping.}
    \label{fig:density-vs-energy}
\end{figure}

\begin{figure}
    \centering
    \includegraphics[width=8.6cm]{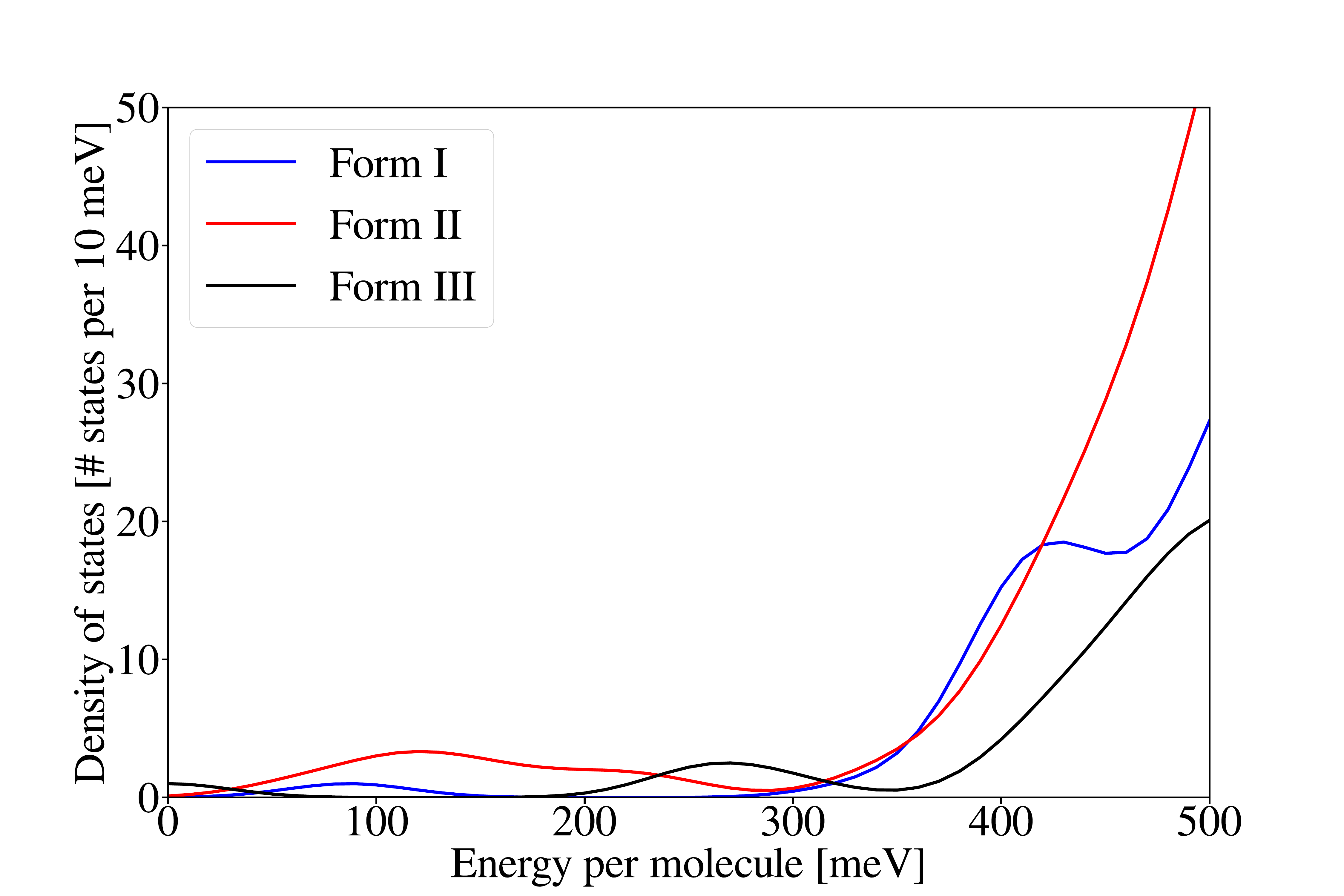}
    \caption{Configurational density of states (DOS) of the different Forms of N-(4’-Methylkbenzylidene)-4-methylalanine. Since higher energy regions were not sampled our DOS goes to zero at high energies.}
    \label{fig:dos}
\end{figure}

\begin{figure}
    \centering
    \includegraphics[width=8.6cm]{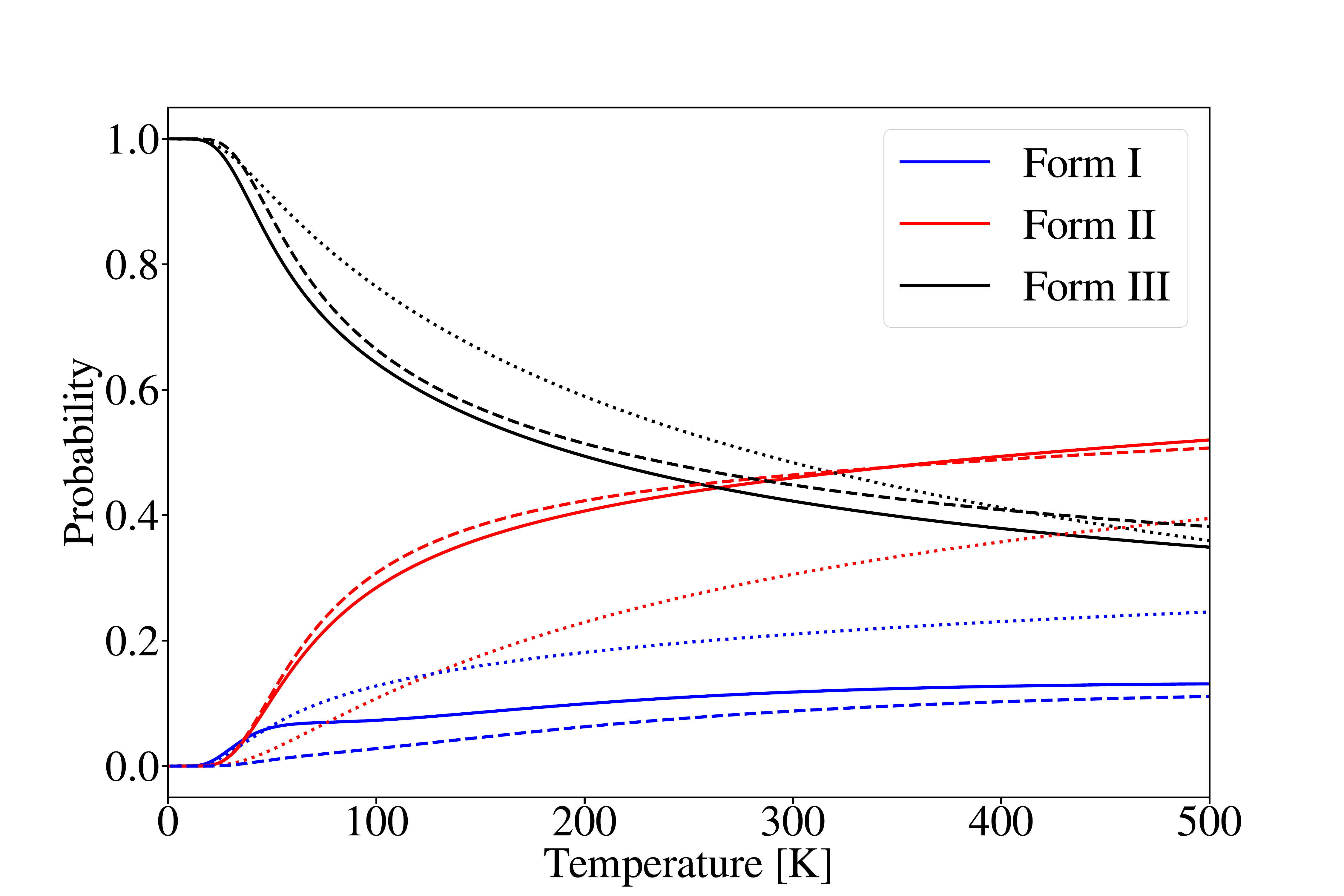}
    \caption{Boltzmann probabilities of finding a certain form of N-(4’-Methylkbenzylidene)-4-methylalanine vs. the temperature. Only taking the lowest polymorphs of each phase is plotted with dotted lines and including all the 300 structures is plotted with solid lines. For checking the convergence, the Boltzmann probabilities of 100 structures is plotted in dashed lines. The fact that there is a crossing of the probabilities between Form II and III only including one structure shows that Form II has a larger vibrational entropy.}
    \label{fig:prob}
\end{figure}

\begin{figure}
    \centering
    \includegraphics[width=8.6cm]{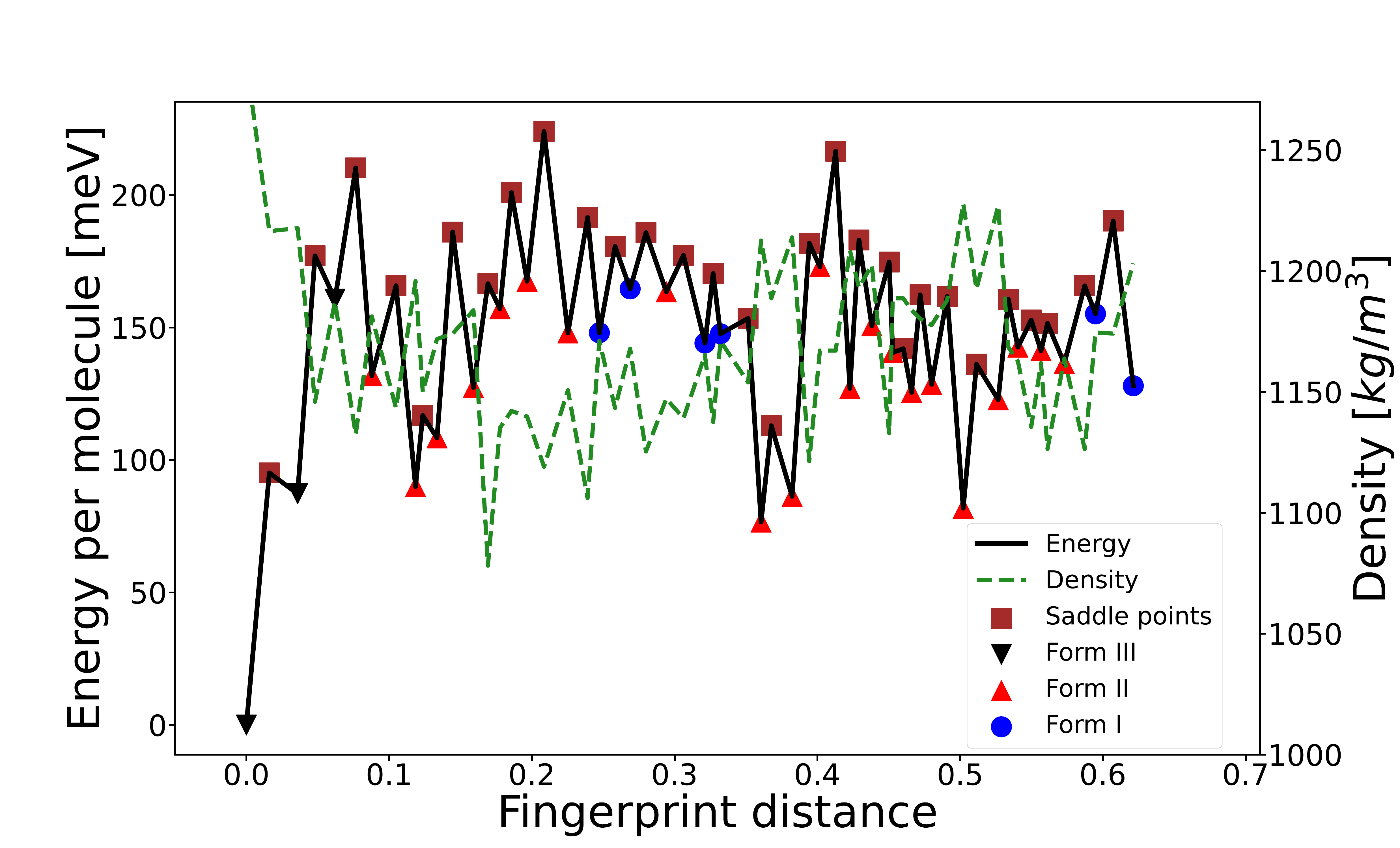}
    \caption{Representative  reaction pathway between structures  of the three known forms of N-(4’-Methylkbenzylidene)-4-methylalanine. The distance between consecutive minima and saddle points along the x-axis is proportional to the fingerprint distance between them calculated with the overlap matrix fingerprint~\cite{zhu2016fingerprint}.}
    \label{fig:path}
\end{figure}

According to this classification, it turns out that structures that belong to Form III have the lowest energy and highest density. Next in energy comes a structure from Form II being about 80 meV  higher in energy than the Form III global minimum. 
The lowest structure belonging to Form I structures is about 100  meV per molecule above the global minimum. 
The energies and densities of the different Forms are shown in Fig.~\ref{fig:density-vs-energy}.
In agreement with numerous other studies~\cite{butler2023reducing, mortazavi2019computational, dybeck2019exploring, price2018control, price2016potential, francia2020systematic}, there is a notable correlation between energy and density, with denser packing corresponding in general to lower energy. \\
Despite Form III having the lowest potential energy, its dense packing renders it structurally rigid, resulting in a few other low-energy minima. Therefore, only 84 structures were found. In contrast, Form I and II are structurally tolerant, i.e, 
there are many related structures that are only slightly higher in energy than the lowest energy structure of this phase. As a result 157 structures were found of Form II and 63 of Form I.
In the case of form II one can for example rotate CH3 groups and slide the layers of the molecules against each other, which is prohibited in Form III due to its dense packing.
As a result the configurational densities of states are much larger for Form I and II as shown in  Fig.~\ref{fig:dos}.
These observations suggest that configurational entropy will stabilize Form I and II compared to Form III at higher temperatures.
However, to get accurate results for the free energy we have to include also the vibrational contribution to the free energy. A combined description of both entropy contributions  can be obtained  with the quantum superposition method~\cite{Walesbook}. 
In this method the partition function $Z$ is the sum of the contributions from all catchment basins $j$.
\begin{equation}
    Z = \sum_{j} \exp(-\beta F_{j}(\beta)) \label{z} ,
\end{equation}
where $\beta = 1/k_{B}T$ and $F_{j}(\beta)$ represents the harmonic free energy of structure $j$.
The total free energy is then obtained in the usual way from $F = -k_B T \ln(Z)$.
The Boltzmann probability $p_{\kappa}(\beta)$ of finding the system in phase $\kappa$ as a function of temperature $T$ is given by the expectation value
\begin{equation}
    p_{\kappa}(\beta) = \frac{\sum_{i \in \kappa} \exp(-\beta F_{i}(\beta))}{\sum_{j} \exp(-\beta F_{j}(\beta))} \label{pkappa} ,
\end{equation}
Whether a structure $i$ belongs to phase $\kappa$ was determined by the two previously described collective variables. 
The method is only valid if on the one hand side the system can visit all low energy catchment basins on an experimental observation timescale and on the other hand transitions from one catchment basin to a neighboring one are rare events. The timings obtained from our MD trajectories, displayed in Table~\ref{tab:md}, show that these two conditions are fulfilled. Crossing from one catchment basin into another belonging to the same phase takes about 10 picoseconds while the vibrational periods are in a range of about 10 femtoseconds to about 1 picosecond.
The Boltzmann probabilities of Eq.~\ref{pkappa} are plotted in Fig.~\ref{fig:prob} against temperature $T = 1/(k_{B} \beta)$ for the case where all sampled structures were included in the partition function (solid lines) and for the case where only the lowest energy structures of each phase was included (dotted lines). The probability obtained by including only the 100 lowest structures is presented with dashed lines.  
The results indicate that the probability of finding Form I is consistently lower than that of Forms II or III, regardless of whether only the lowest energy structures or all sampled structures are considered. 
Notably, when only the two lowest structures of Forms II and III are included, a phase transition is observed at 470~K, compared to 270K when all 300 structures are considered resulting in a error of 200K. 
Calculating the Boltzmann probabilities for only the 100 lowest structures results in the more or less the same crossing temperature  (290 K)  which implies that the configurational space is sampled well enough to reach convergence.
This means that while Form III is the lowest in potential energy, Form II's structural tolerance results in a lower free energy at elevated temperatures.
The much larger configurational entropy of Form I and II became also visible in the behavior of MD 
trajectories at 450~K. Within about 500 picoseconds we could observe transitions from Form III to Forms II,  but never a reverse transition. 
To get a better understanding of the potential energy landscape of our molecular crystal we also calculated the saddle points along a Minima Hopping pathway~\cite{schaefer2014minima} on the DFT level. 
The exact saddle points between consecutive minima were found 
with the compass method~\cite{sommer2024compass}.
A representative reaction pathway, illustrated in Fig.~\ref{fig:path}, shows that transitions within different minima of Form III require the largest amount of energy, with several intermediate minima being visited on the path to Form II. 
The highest barrier is approximately 250 meV per molecule. Transitions among Form I and Form II structures as well as transitions between Forms I and II structures  cross even lower barriers of  10 to 100 meV per molecule. 
Although Form I appears in this pathway, Form II is generally lower in potential energy and more frequently encountered. 
The fact that our MD trajectories never made transitions from Form I and II into Form III in spite of the fact that the energy barriers are not very high suggests that the transformations between phases are slowed down by entropy effects.
We have shown that in a molecular crystal consisting of N-(4’-Methylbenzylidene)-4-methylalanine molecules, different phases have vastly different configurational densities of states. These differences must not be neglected in the calculation of free energies and the associated temperatures for solid-solid phase transitions. Neglecting them for our relatively simple molecular crystal 
leads already to errors of about 200 K. For more complex molecular crystals the effects of configurational entropy are presumably even larger. Phases that are less closely packed are expected to have in general higher configurational entropies than the closely packed structures that typically give the lowest energies. In addition, the lower density phases  have in general lower vibrational frequencies which results in a larger vibrational entropy and consequently also stabilizes these phases.
Hence, structurally tolerant low density phases are usually thermodynamically stabilized. Accurate free energies and Boltzmann probabilities, that include both effects, can be obtained with the quantum superposition method based on machine learned potential energy surfaces.

\begin{table}
\centering
\begin{tabular}{|c|c|c|c|} \hline
  Temperature/Form  & I & II & III  \\ \hline
 300 K & 7 (9)    & 14 (105)  &  5 (73)   \\
 350 K & 11 (17)  & 29 (199)  &  22 (114) \\
 400 K & 22 (128) & 50 (203)  &  10 (185) \\ \hline
\end{tabular}
\caption{Number of distinct minima found along an NVT MD trajectory. During a 300 ps MD trajectory, 
calculated with LAMMPS~\cite{thompson2022lammps}. A local geometry optimization was performed with the SQNM method~\cite{gubler2023efficient} after each elapsed ps. 
The relaxed structures were then analyzed using overlap matrix fingerprinting~\cite{zhu2016fingerprint} to distinguish different minima.
In brackets is the number of hops between catchment basins during the MD is shown. The ration between the number inside and before the brackets gives an indication of how many times certain catchment basins were visited. 
The starting point of the MD was the lowest energy structure.
}
\label{tab:md}
\end{table}

\subsection*{ACKNOWLEDGMENTS}
The work was supported by the Swiss National Science Foundation (SNSF) under project 200020E\_206936 and 200021\_191994.
Computing time was provided by the Swiss National Supercomputing Centre (CSCS) under project ID s1167 and lp08 as well as the Swiss share of the LUMI system under project ID 465000731. Calculations were also performed at sciCORE (http://scicore.unibas.ch/), the scientific computing center of the  University of Basel. G.F. acknowledges the support from the Italian National Centre of HPC, Big Data, and Quantum Computing (grant number CN00000013), funded through the Next Generation EU initiative.

\pagebreak
\bibliography{sym_bias}

\begin{thebibliography}{53}%
\makeatletter
\providecommand \@ifxundefined [1]{%
 \@ifx{#1\undefined}
}%
\providecommand \@ifnum [1]{%
 \ifnum #1\expandafter \@firstoftwo
 \else \expandafter \@secondoftwo
 \fi
}%
\providecommand \@ifx [1]{%
 \ifx #1\expandafter \@firstoftwo
 \else \expandafter \@secondoftwo
 \fi
}%
\providecommand \natexlab [1]{#1}%
\providecommand \enquote  [1]{``#1''}%
\providecommand \bibnamefont  [1]{#1}%
\providecommand \bibfnamefont [1]{#1}%
\providecommand \citenamefont [1]{#1}%
\providecommand \href@noop [0]{\@secondoftwo}%
\providecommand \href [0]{\begingroup \@sanitize@url \@href}%
\providecommand \@href[1]{\@@startlink{#1}\@@href}%
\providecommand \@@href[1]{\endgroup#1\@@endlink}%
\providecommand \@sanitize@url [0]{\catcode `\\12\catcode `\$12\catcode `\&12\catcode `\#12\catcode `\^12\catcode `\_12\catcode `\%12\relax}%
\providecommand \@@startlink[1]{}%
\providecommand \@@endlink[0]{}%
\providecommand \url  [0]{\begingroup\@sanitize@url \@url }%
\providecommand \@url [1]{\endgroup\@href {#1}{\urlprefix }}%
\providecommand \urlprefix  [0]{URL }%
\providecommand \Eprint [0]{\href }%
\providecommand \doibase [0]{https://doi.org/}%
\providecommand \selectlanguage [0]{\@gobble}%
\providecommand \bibinfo  [0]{\@secondoftwo}%
\providecommand \bibfield  [0]{\@secondoftwo}%
\providecommand \translation [1]{[#1]}%
\providecommand \BibitemOpen [0]{}%
\providecommand \bibitemStop [0]{}%
\providecommand \bibitemNoStop [0]{.\EOS\space}%
\providecommand \EOS [0]{\spacefactor3000\relax}%
\providecommand \BibitemShut  [1]{\csname bibitem#1\endcsname}%
\let\auto@bib@innerbib\@empty
\bibitem [{\citenamefont {Cruz-Cabeza}\ \emph {et~al.}(2015)\citenamefont {Cruz-Cabeza}, \citenamefont {Reutzel-Edens},\ and\ \citenamefont {Bernstein}}]{cruz2015facts}%
  \BibitemOpen
  \bibfield  {author} {\bibinfo {author} {\bibfnamefont {A.~J.}\ \bibnamefont {Cruz-Cabeza}}, \bibinfo {author} {\bibfnamefont {S.~M.}\ \bibnamefont {Reutzel-Edens}},\ and\ \bibinfo {author} {\bibfnamefont {J.}~\bibnamefont {Bernstein}},\ }\bibfield  {title} {\bibinfo {title} {Facts and fictions about polymorphism},\ }\href {https://doi.org/https://doi.org/10.1039/C5CS00227C} {\bibfield  {journal} {\bibinfo  {journal} {Chem. Soc. Rev.}\ }\textbf {\bibinfo {volume} {44}},\ \bibinfo {pages} {8619} (\bibinfo {year} {2015})}\BibitemShut {NoStop}%
\bibitem [{\citenamefont {Zhu}\ \emph {et~al.}(2022)\citenamefont {Zhu}, \citenamefont {Johal}, \citenamefont {Widdowson}, \citenamefont {Pang}, \citenamefont {Li}, \citenamefont {Kane}, \citenamefont {Kurlin}, \citenamefont {Day}, \citenamefont {Little},\ and\ \citenamefont {Cooper}}]{zhu2022analogy}%
  \BibitemOpen
  \bibfield  {author} {\bibinfo {author} {\bibfnamefont {Q.}~\bibnamefont {Zhu}}, \bibinfo {author} {\bibfnamefont {J.}~\bibnamefont {Johal}}, \bibinfo {author} {\bibfnamefont {D.~E.}\ \bibnamefont {Widdowson}}, \bibinfo {author} {\bibfnamefont {Z.}~\bibnamefont {Pang}}, \bibinfo {author} {\bibfnamefont {B.}~\bibnamefont {Li}}, \bibinfo {author} {\bibfnamefont {C.~M.}\ \bibnamefont {Kane}}, \bibinfo {author} {\bibfnamefont {V.}~\bibnamefont {Kurlin}}, \bibinfo {author} {\bibfnamefont {G.~M.}\ \bibnamefont {Day}}, \bibinfo {author} {\bibfnamefont {M.~A.}\ \bibnamefont {Little}},\ and\ \bibinfo {author} {\bibfnamefont {A.~I.}\ \bibnamefont {Cooper}},\ }\bibfield  {title} {\bibinfo {title} {Analogy powered by prediction and structural invariants: computationally led discovery of a mesoporous hydrogen-bonded organic cage crystal},\ }\href {https://doi.org/https://doi.org/10.1021/jacs.2c02653} {\bibfield  {journal} {\bibinfo  {journal} {J. Am. Chem. Soc.}\ }\textbf {\bibinfo {volume} {144}},\ \bibinfo {pages}
  {9893} (\bibinfo {year} {2022})}\BibitemShut {NoStop}%
\bibitem [{\citenamefont {Cui}\ \emph {et~al.}(2019)\citenamefont {Cui}, \citenamefont {McMahon}, \citenamefont {Spackman}, \citenamefont {Alston}, \citenamefont {Little}, \citenamefont {Day},\ and\ \citenamefont {Cooper}}]{cui2019mining}%
  \BibitemOpen
  \bibfield  {author} {\bibinfo {author} {\bibfnamefont {P.}~\bibnamefont {Cui}}, \bibinfo {author} {\bibfnamefont {D.~P.}\ \bibnamefont {McMahon}}, \bibinfo {author} {\bibfnamefont {P.~R.}\ \bibnamefont {Spackman}}, \bibinfo {author} {\bibfnamefont {B.~M.}\ \bibnamefont {Alston}}, \bibinfo {author} {\bibfnamefont {M.~A.}\ \bibnamefont {Little}}, \bibinfo {author} {\bibfnamefont {G.~M.}\ \bibnamefont {Day}},\ and\ \bibinfo {author} {\bibfnamefont {A.~I.}\ \bibnamefont {Cooper}},\ }\bibfield  {title} {\bibinfo {title} {Mining predicted crystal structure landscapes with high throughput crystallisation: old molecules, new insights},\ }\href {https://doi.org/https://doi.org/10.1039/C9SC02832C} {\bibfield  {journal} {\bibinfo  {journal} {Chemical Science}\ }\textbf {\bibinfo {volume} {10}},\ \bibinfo {pages} {9988} (\bibinfo {year} {2019})}\BibitemShut {NoStop}%
\bibitem [{\citenamefont {Neumann}\ \emph {et~al.}(2015)\citenamefont {Neumann}, \citenamefont {Van De~Streek}, \citenamefont {Fabbiani}, \citenamefont {Hidber},\ and\ \citenamefont {Grassmann}}]{neumann2015combined}%
  \BibitemOpen
  \bibfield  {author} {\bibinfo {author} {\bibfnamefont {M.}~\bibnamefont {Neumann}}, \bibinfo {author} {\bibfnamefont {J.}~\bibnamefont {Van De~Streek}}, \bibinfo {author} {\bibfnamefont {F.}~\bibnamefont {Fabbiani}}, \bibinfo {author} {\bibfnamefont {P.}~\bibnamefont {Hidber}},\ and\ \bibinfo {author} {\bibfnamefont {O.}~\bibnamefont {Grassmann}},\ }\bibfield  {title} {\bibinfo {title} {Combined crystal structure prediction and high-pressure crystallization in rational pharmaceutical polymorph screening},\ }\href {https://doi.org/https://doi.org/10.1038/ncomms8793} {\bibfield  {journal} {\bibinfo  {journal} {Nature communications}\ }\textbf {\bibinfo {volume} {6}},\ \bibinfo {pages} {7793} (\bibinfo {year} {2015})}\BibitemShut {NoStop}%
\bibitem [{\citenamefont {Shtukenberg}\ \emph {et~al.}(2017)\citenamefont {Shtukenberg}, \citenamefont {Zhu}, \citenamefont {Carter}, \citenamefont {Vogt}, \citenamefont {Hoja}, \citenamefont {Schneider}, \citenamefont {Song}, \citenamefont {Pokroy}, \citenamefont {Polishchuk}, \citenamefont {Tkatchenko} \emph {et~al.}}]{shtukenberg2017powder}%
  \BibitemOpen
  \bibfield  {author} {\bibinfo {author} {\bibfnamefont {A.~G.}\ \bibnamefont {Shtukenberg}}, \bibinfo {author} {\bibfnamefont {Q.}~\bibnamefont {Zhu}}, \bibinfo {author} {\bibfnamefont {D.~J.}\ \bibnamefont {Carter}}, \bibinfo {author} {\bibfnamefont {L.}~\bibnamefont {Vogt}}, \bibinfo {author} {\bibfnamefont {J.}~\bibnamefont {Hoja}}, \bibinfo {author} {\bibfnamefont {E.}~\bibnamefont {Schneider}}, \bibinfo {author} {\bibfnamefont {H.}~\bibnamefont {Song}}, \bibinfo {author} {\bibfnamefont {B.}~\bibnamefont {Pokroy}}, \bibinfo {author} {\bibfnamefont {I.}~\bibnamefont {Polishchuk}}, \bibinfo {author} {\bibfnamefont {A.}~\bibnamefont {Tkatchenko}}, \emph {et~al.},\ }\bibfield  {title} {\bibinfo {title} {Powder diffraction and crystal structure prediction identify four new coumarin polymorphs},\ }\href {https://doi.org/https://doi.org/10.1039/C7SC00168A} {\bibfield  {journal} {\bibinfo  {journal} {Chemical Science}\ }\textbf {\bibinfo {volume} {8}},\ \bibinfo {pages} {4926} (\bibinfo {year}
  {2017})}\BibitemShut {NoStop}%
\bibitem [{\citenamefont {Braun}\ \emph {et~al.}(2011)\citenamefont {Braun}, \citenamefont {Ardid-Candel}, \citenamefont {D’Oria}, \citenamefont {Karamertzanis}, \citenamefont {Arlin}, \citenamefont {Florence}, \citenamefont {Jones},\ and\ \citenamefont {Price}}]{braun2011racemic}%
  \BibitemOpen
  \bibfield  {author} {\bibinfo {author} {\bibfnamefont {D.~E.}\ \bibnamefont {Braun}}, \bibinfo {author} {\bibfnamefont {M.}~\bibnamefont {Ardid-Candel}}, \bibinfo {author} {\bibfnamefont {E.}~\bibnamefont {D’Oria}}, \bibinfo {author} {\bibfnamefont {P.~G.}\ \bibnamefont {Karamertzanis}}, \bibinfo {author} {\bibfnamefont {J.-B.}\ \bibnamefont {Arlin}}, \bibinfo {author} {\bibfnamefont {A.~J.}\ \bibnamefont {Florence}}, \bibinfo {author} {\bibfnamefont {A.~G.}\ \bibnamefont {Jones}},\ and\ \bibinfo {author} {\bibfnamefont {S.~L.}\ \bibnamefont {Price}},\ }\bibfield  {title} {\bibinfo {title} {Racemic naproxen: a multidisciplinary structural and thermodynamic comparison with the enantiopure form},\ }\href {https://doi.org/https://doi.org/10.1021/cg201203u} {\bibfield  {journal} {\bibinfo  {journal} {Crystal growth \& design}\ }\textbf {\bibinfo {volume} {11}},\ \bibinfo {pages} {5659} (\bibinfo {year} {2011})}\BibitemShut {NoStop}%
\bibitem [{\citenamefont {Day}(2011)}]{day2011current}%
  \BibitemOpen
  \bibfield  {author} {\bibinfo {author} {\bibfnamefont {G.~M.}\ \bibnamefont {Day}},\ }\bibfield  {title} {\bibinfo {title} {Current approaches to predicting molecular organic crystal structures},\ }\href {https://doi.org/https://doi.org/10.1080/0889311X.2010.517526} {\bibfield  {journal} {\bibinfo  {journal} {Crystallography Reviews}\ }\textbf {\bibinfo {volume} {17}},\ \bibinfo {pages} {3} (\bibinfo {year} {2011})}\BibitemShut {NoStop}%
\bibitem [{\citenamefont {Nyman}\ and\ \citenamefont {Reutzel-Edens}(2018)}]{nyman2018crystal}%
  \BibitemOpen
  \bibfield  {author} {\bibinfo {author} {\bibfnamefont {J.}~\bibnamefont {Nyman}}\ and\ \bibinfo {author} {\bibfnamefont {S.~M.}\ \bibnamefont {Reutzel-Edens}},\ }\bibfield  {title} {\bibinfo {title} {Crystal structure prediction is changing from basic science to applied technology},\ }\href {https://doi.org/https://doi.org/10.1039/C8FD00033F} {\bibfield  {journal} {\bibinfo  {journal} {Faraday Discussions}\ }\textbf {\bibinfo {volume} {211}},\ \bibinfo {pages} {459} (\bibinfo {year} {2018})}\BibitemShut {NoStop}%
\bibitem [{\citenamefont {Thakur}\ \emph {et~al.}(2015)\citenamefont {Thakur}, \citenamefont {Dubey},\ and\ \citenamefont {Desiraju}}]{thakur2015crystal}%
  \BibitemOpen
  \bibfield  {author} {\bibinfo {author} {\bibfnamefont {T.~S.}\ \bibnamefont {Thakur}}, \bibinfo {author} {\bibfnamefont {R.}~\bibnamefont {Dubey}},\ and\ \bibinfo {author} {\bibfnamefont {G.~R.}\ \bibnamefont {Desiraju}},\ }\bibfield  {title} {\bibinfo {title} {Crystal structure and prediction},\ }\href {https://doi.org/https://doi.org/10.1146/annurev-physchem-040214-121452} {\bibfield  {journal} {\bibinfo  {journal} {Annu. Rev. Phys. Chem.}\ }\textbf {\bibinfo {volume} {66}},\ \bibinfo {pages} {21} (\bibinfo {year} {2015})}\BibitemShut {NoStop}%
\bibitem [{\citenamefont {Mortazavi}\ \emph {et~al.}(2019)\citenamefont {Mortazavi}, \citenamefont {Hoja}, \citenamefont {Aerts}, \citenamefont {Qu{\'e}r{\'e}}, \citenamefont {van~de Streek}, \citenamefont {Neumann},\ and\ \citenamefont {Tkatchenko}}]{mortazavi2019computational}%
  \BibitemOpen
  \bibfield  {author} {\bibinfo {author} {\bibfnamefont {M.}~\bibnamefont {Mortazavi}}, \bibinfo {author} {\bibfnamefont {J.}~\bibnamefont {Hoja}}, \bibinfo {author} {\bibfnamefont {L.}~\bibnamefont {Aerts}}, \bibinfo {author} {\bibfnamefont {L.}~\bibnamefont {Qu{\'e}r{\'e}}}, \bibinfo {author} {\bibfnamefont {J.}~\bibnamefont {van~de Streek}}, \bibinfo {author} {\bibfnamefont {M.~A.}\ \bibnamefont {Neumann}},\ and\ \bibinfo {author} {\bibfnamefont {A.}~\bibnamefont {Tkatchenko}},\ }\bibfield  {title} {\bibinfo {title} {Computational polymorph screening reveals late-appearing and poorly-soluble form of rotigotine},\ }\href {https://doi.org/https://doi.org/10.1038/s42004-019-0171-y} {\bibfield  {journal} {\bibinfo  {journal} {Communications Chemistry}\ }\textbf {\bibinfo {volume} {2}},\ \bibinfo {pages} {70} (\bibinfo {year} {2019})}\BibitemShut {NoStop}%
\bibitem [{\citenamefont {Habgood}(2011)}]{habgood2011form}%
  \BibitemOpen
  \bibfield  {author} {\bibinfo {author} {\bibfnamefont {M.}~\bibnamefont {Habgood}},\ }\bibfield  {title} {\bibinfo {title} {Form ii caffeine: a case study for confirming and predicting disorder in organic crystals},\ }\href@noop {} {\bibfield  {journal} {\bibinfo  {journal} {Crystal growth \& design}\ }\textbf {\bibinfo {volume} {11}},\ \bibinfo {pages} {3600} (\bibinfo {year} {2011})}\BibitemShut {NoStop}%
\bibitem [{\citenamefont {Stevenson}\ \emph {et~al.}(2019)\citenamefont {Stevenson}, \citenamefont {Lancaster}, \citenamefont {Buanz}, \citenamefont {Price}, \citenamefont {Tocher},\ and\ \citenamefont {Price}}]{stevenson2019solid}%
  \BibitemOpen
  \bibfield  {author} {\bibinfo {author} {\bibfnamefont {E.~L.}\ \bibnamefont {Stevenson}}, \bibinfo {author} {\bibfnamefont {R.~W.}\ \bibnamefont {Lancaster}}, \bibinfo {author} {\bibfnamefont {A.~B.}\ \bibnamefont {Buanz}}, \bibinfo {author} {\bibfnamefont {L.~S.}\ \bibnamefont {Price}}, \bibinfo {author} {\bibfnamefont {D.~A.}\ \bibnamefont {Tocher}},\ and\ \bibinfo {author} {\bibfnamefont {S.~L.}\ \bibnamefont {Price}},\ }\bibfield  {title} {\bibinfo {title} {The solid state forms of the sex hormone 17-$\beta$-estradiol},\ }\href@noop {} {\bibfield  {journal} {\bibinfo  {journal} {CrystEngComm}\ }\textbf {\bibinfo {volume} {21}},\ \bibinfo {pages} {2154} (\bibinfo {year} {2019})}\BibitemShut {NoStop}%
\bibitem [{\citenamefont {Case}\ \emph {et~al.}(2016)\citenamefont {Case}, \citenamefont {Campbell}, \citenamefont {Bygrave},\ and\ \citenamefont {Day}}]{case2016convergence}%
  \BibitemOpen
  \bibfield  {author} {\bibinfo {author} {\bibfnamefont {D.~H.}\ \bibnamefont {Case}}, \bibinfo {author} {\bibfnamefont {J.~E.}\ \bibnamefont {Campbell}}, \bibinfo {author} {\bibfnamefont {P.~J.}\ \bibnamefont {Bygrave}},\ and\ \bibinfo {author} {\bibfnamefont {G.~M.}\ \bibnamefont {Day}},\ }\bibfield  {title} {\bibinfo {title} {Convergence properties of crystal structure prediction by quasi-random sampling},\ }\href {https://doi.org/https://doi.org/10.1021/acs.jctc.5b01112} {\bibfield  {journal} {\bibinfo  {journal} {J. Chem. Theory Comput.}\ }\textbf {\bibinfo {volume} {12}},\ \bibinfo {pages} {910} (\bibinfo {year} {2016})}\BibitemShut {NoStop}%
\bibitem [{\citenamefont {Neumann}\ \emph {et~al.}(2008)\citenamefont {Neumann}, \citenamefont {Leusen},\ and\ \citenamefont {Kendrick}}]{neumann2008major}%
  \BibitemOpen
  \bibfield  {author} {\bibinfo {author} {\bibfnamefont {M.~A.}\ \bibnamefont {Neumann}}, \bibinfo {author} {\bibfnamefont {F.~J.}\ \bibnamefont {Leusen}},\ and\ \bibinfo {author} {\bibfnamefont {J.}~\bibnamefont {Kendrick}},\ }\bibfield  {title} {\bibinfo {title} {A major advance in crystal structure prediction},\ }\href {https://doi.org/https://doi.org/10.1002/anie.200704247} {\bibfield  {journal} {\bibinfo  {journal} {Angewandte Chemie International Edition}\ }\textbf {\bibinfo {volume} {47}},\ \bibinfo {pages} {2427} (\bibinfo {year} {2008})}\BibitemShut {NoStop}%
\bibitem [{\citenamefont {Kronik}\ and\ \citenamefont {Tkatchenko}(2014)}]{kronik2014understanding}%
  \BibitemOpen
  \bibfield  {author} {\bibinfo {author} {\bibfnamefont {L.}~\bibnamefont {Kronik}}\ and\ \bibinfo {author} {\bibfnamefont {A.}~\bibnamefont {Tkatchenko}},\ }\bibfield  {title} {\bibinfo {title} {Understanding molecular crystals with dispersion-inclusive density functional theory: pairwise corrections and beyond},\ }\href {https://doi.org/https://doi.org/10.1021/ar500144s} {\bibfield  {journal} {\bibinfo  {journal} {Acc. Chem. Res.}\ }\textbf {\bibinfo {volume} {47}},\ \bibinfo {pages} {3208} (\bibinfo {year} {2014})}\BibitemShut {NoStop}%
\bibitem [{\citenamefont {Hunnisett}\ \emph {et~al.}(2024)\citenamefont {Hunnisett}, \citenamefont {Nyman}, \citenamefont {Francia}, \citenamefont {Abraham}, \citenamefont {Adjiman}, \citenamefont {Aitipamula}, \citenamefont {Alkhidir}, \citenamefont {Almehairbi}, \citenamefont {Anelli}, \citenamefont {Anstine} \emph {et~al.}}]{hunnisett2024seventh}%
  \BibitemOpen
  \bibfield  {author} {\bibinfo {author} {\bibfnamefont {L.~M.}\ \bibnamefont {Hunnisett}}, \bibinfo {author} {\bibfnamefont {J.}~\bibnamefont {Nyman}}, \bibinfo {author} {\bibfnamefont {N.}~\bibnamefont {Francia}}, \bibinfo {author} {\bibfnamefont {N.~S.}\ \bibnamefont {Abraham}}, \bibinfo {author} {\bibfnamefont {C.~S.}\ \bibnamefont {Adjiman}}, \bibinfo {author} {\bibfnamefont {S.}~\bibnamefont {Aitipamula}}, \bibinfo {author} {\bibfnamefont {T.}~\bibnamefont {Alkhidir}}, \bibinfo {author} {\bibfnamefont {M.}~\bibnamefont {Almehairbi}}, \bibinfo {author} {\bibfnamefont {A.}~\bibnamefont {Anelli}}, \bibinfo {author} {\bibfnamefont {D.~M.}\ \bibnamefont {Anstine}}, \emph {et~al.},\ }\bibfield  {title} {\bibinfo {title} {The seventh blind test of crystal structure prediction: structure generation methods},\ }\bibfield  {journal} {\bibinfo  {journal} {Structural Science}\ }\textbf {\bibinfo {volume} {80}},\ \href {https://doi.org/https://doi.org/10.1107/S2052520624007492}
  {https://doi.org/10.1107/S2052520624007492} (\bibinfo {year} {2024})\BibitemShut {NoStop}%
\bibitem [{\citenamefont {Dybeck}\ \emph {et~al.}(2019)\citenamefont {Dybeck}, \citenamefont {McMahon}, \citenamefont {Day},\ and\ \citenamefont {Shirts}}]{dybeck2019exploring}%
  \BibitemOpen
  \bibfield  {author} {\bibinfo {author} {\bibfnamefont {E.~C.}\ \bibnamefont {Dybeck}}, \bibinfo {author} {\bibfnamefont {D.~P.}\ \bibnamefont {McMahon}}, \bibinfo {author} {\bibfnamefont {G.~M.}\ \bibnamefont {Day}},\ and\ \bibinfo {author} {\bibfnamefont {M.~R.}\ \bibnamefont {Shirts}},\ }\bibfield  {title} {\bibinfo {title} {Exploring the multi-minima behavior of small molecule crystal polymorphs at finite temperature},\ }\href {https://doi.org/https://doi.org/10.1021/acs.cgd.9b00476} {\bibfield  {journal} {\bibinfo  {journal} {Crystal Growth \& Design}\ }\textbf {\bibinfo {volume} {19}},\ \bibinfo {pages} {5568} (\bibinfo {year} {2019})}\BibitemShut {NoStop}%
\bibitem [{\citenamefont {Sch{\"o}n}\ \emph {et~al.}(1996)\citenamefont {Sch{\"o}n}, \citenamefont {Putz},\ and\ \citenamefont {Jansen}}]{schon1996studying}%
  \BibitemOpen
  \bibfield  {author} {\bibinfo {author} {\bibfnamefont {J.}~\bibnamefont {Sch{\"o}n}}, \bibinfo {author} {\bibfnamefont {H.}~\bibnamefont {Putz}},\ and\ \bibinfo {author} {\bibfnamefont {M.}~\bibnamefont {Jansen}},\ }\bibfield  {title} {\bibinfo {title} {Studying the energy hypersurface of continuous systems-the threshold algorithm},\ }\href {https://doi.org/10.1088/0953-8984/8/2/004} {\bibfield  {journal} {\bibinfo  {journal} {J. Condens. Matter Phys}\ }\textbf {\bibinfo {volume} {8}},\ \bibinfo {pages} {143} (\bibinfo {year} {1996})}\BibitemShut {NoStop}%
\bibitem [{\citenamefont {Butler}\ and\ \citenamefont {Day}(2023)}]{butler2023reducing}%
  \BibitemOpen
  \bibfield  {author} {\bibinfo {author} {\bibfnamefont {P.~W.}\ \bibnamefont {Butler}}\ and\ \bibinfo {author} {\bibfnamefont {G.~M.}\ \bibnamefont {Day}},\ }\bibfield  {title} {\bibinfo {title} {Reducing overprediction of molecular crystal structures via threshold clustering},\ }\href {https://doi.org/https://doi.org/10.1073/pnas.2300516120} {\bibfield  {journal} {\bibinfo  {journal} {Proceedings of the National Academy of Sciences}\ }\textbf {\bibinfo {volume} {120}},\ \bibinfo {pages} {e2300516120} (\bibinfo {year} {2023})}\BibitemShut {NoStop}%
\bibitem [{\citenamefont {Yang}\ and\ \citenamefont {Day}(2022)}]{yang2022global}%
  \BibitemOpen
  \bibfield  {author} {\bibinfo {author} {\bibfnamefont {S.}~\bibnamefont {Yang}}\ and\ \bibinfo {author} {\bibfnamefont {G.~M.}\ \bibnamefont {Day}},\ }\bibfield  {title} {\bibinfo {title} {Global analysis of the energy landscapes of molecular crystal structures by applying the threshold algorithm},\ }\href {https://doi.org/https://doi.org/10.1038/s42004-022-00705-4} {\bibfield  {journal} {\bibinfo  {journal} {Commun. Chem.}\ }\textbf {\bibinfo {volume} {5}},\ \bibinfo {pages} {86} (\bibinfo {year} {2022})}\BibitemShut {NoStop}%
\bibitem [{\citenamefont {Montis}\ \emph {et~al.}(2021)\citenamefont {Montis}, \citenamefont {Hursthouse}, \citenamefont {Kendrick}, \citenamefont {Howe},\ and\ \citenamefont {Whitby}}]{montis2021combining}%
  \BibitemOpen
  \bibfield  {author} {\bibinfo {author} {\bibfnamefont {R.}~\bibnamefont {Montis}}, \bibinfo {author} {\bibfnamefont {M.~B.}\ \bibnamefont {Hursthouse}}, \bibinfo {author} {\bibfnamefont {J.}~\bibnamefont {Kendrick}}, \bibinfo {author} {\bibfnamefont {J.}~\bibnamefont {Howe}},\ and\ \bibinfo {author} {\bibfnamefont {R.~J.}\ \bibnamefont {Whitby}},\ }\bibfield  {title} {\bibinfo {title} {Combining structural rugosity and crystal packing comparison: A route to more polymorphs?},\ }\href {https://doi.org/https://doi.org/10.1021/acs.cgd.1c01132} {\bibfield  {journal} {\bibinfo  {journal} {Crystal Growth \& Design}\ }\textbf {\bibinfo {volume} {22}},\ \bibinfo {pages} {559} (\bibinfo {year} {2021})}\BibitemShut {NoStop}%
\bibitem [{\citenamefont {Beyer}\ \emph {et~al.}(2001)\citenamefont {Beyer}, \citenamefont {Day},\ and\ \citenamefont {Price}}]{beyer2001prediction}%
  \BibitemOpen
  \bibfield  {author} {\bibinfo {author} {\bibfnamefont {T.}~\bibnamefont {Beyer}}, \bibinfo {author} {\bibfnamefont {G.~M.}\ \bibnamefont {Day}},\ and\ \bibinfo {author} {\bibfnamefont {S.~L.}\ \bibnamefont {Price}},\ }\bibfield  {title} {\bibinfo {title} {The prediction, morphology, and mechanical properties of the polymorphs of paracetamol},\ }\href {https://doi.org/https://doi.org/10.1021/ja0102787} {\bibfield  {journal} {\bibinfo  {journal} {J. Am. Chem. Soc.}\ }\textbf {\bibinfo {volume} {123}},\ \bibinfo {pages} {5086} (\bibinfo {year} {2001})}\BibitemShut {NoStop}%
\bibitem [{\citenamefont {Montis}\ \emph {et~al.}(2020)\citenamefont {Montis}, \citenamefont {Davey}, \citenamefont {Wright}, \citenamefont {Woollam},\ and\ \citenamefont {Cruz-Cabeza}}]{montis2020transforming}%
  \BibitemOpen
  \bibfield  {author} {\bibinfo {author} {\bibfnamefont {R.}~\bibnamefont {Montis}}, \bibinfo {author} {\bibfnamefont {R.~J.}\ \bibnamefont {Davey}}, \bibinfo {author} {\bibfnamefont {S.~E.}\ \bibnamefont {Wright}}, \bibinfo {author} {\bibfnamefont {G.~R.}\ \bibnamefont {Woollam}},\ and\ \bibinfo {author} {\bibfnamefont {A.~J.}\ \bibnamefont {Cruz-Cabeza}},\ }\bibfield  {title} {\bibinfo {title} {Transforming computed energy landscapes into experimental realities: the role of structural rugosity},\ }\href {https://doi.org/https://doi.org/10.1002/ange.202006939} {\bibfield  {journal} {\bibinfo  {journal} {Angewandte Chemie}\ }\textbf {\bibinfo {volume} {132}},\ \bibinfo {pages} {20537} (\bibinfo {year} {2020})}\BibitemShut {NoStop}%
\bibitem [{\citenamefont {Francia}\ \emph {et~al.}(2020)\citenamefont {Francia}, \citenamefont {Price}, \citenamefont {Nyman}, \citenamefont {Price},\ and\ \citenamefont {Salvalaglio}}]{francia2020systematic}%
  \BibitemOpen
  \bibfield  {author} {\bibinfo {author} {\bibfnamefont {N.~F.}\ \bibnamefont {Francia}}, \bibinfo {author} {\bibfnamefont {L.~S.}\ \bibnamefont {Price}}, \bibinfo {author} {\bibfnamefont {J.}~\bibnamefont {Nyman}}, \bibinfo {author} {\bibfnamefont {S.~L.}\ \bibnamefont {Price}},\ and\ \bibinfo {author} {\bibfnamefont {M.}~\bibnamefont {Salvalaglio}},\ }\bibfield  {title} {\bibinfo {title} {Systematic finite-temperature reduction of crystal energy landscapes},\ }\href {https://doi.org/https://doi.org/10.1021/acs.cgd.0c00918} {\bibfield  {journal} {\bibinfo  {journal} {Crystal Growth \& Design}\ }\textbf {\bibinfo {volume} {20}},\ \bibinfo {pages} {6847} (\bibinfo {year} {2020})}\BibitemShut {NoStop}%
\bibitem [{\citenamefont {Schneider}\ \emph {et~al.}(2016)\citenamefont {Schneider}, \citenamefont {Vogt},\ and\ \citenamefont {Tuckerman}}]{schneider2016exploring}%
  \BibitemOpen
  \bibfield  {author} {\bibinfo {author} {\bibfnamefont {E.}~\bibnamefont {Schneider}}, \bibinfo {author} {\bibfnamefont {L.}~\bibnamefont {Vogt}},\ and\ \bibinfo {author} {\bibfnamefont {M.~E.}\ \bibnamefont {Tuckerman}},\ }\bibfield  {title} {\bibinfo {title} {Exploring polymorphism of benzene and naphthalene with free energy based enhanced molecular dynamics},\ }\href {https://doi.org/https://doi.org/10.1107/S2052520616007873} {\bibfield  {journal} {\bibinfo  {journal} {Acta Crystallographica Section B: Structural Science, Crystal Engineering and Materials}\ }\textbf {\bibinfo {volume} {72}},\ \bibinfo {pages} {542} (\bibinfo {year} {2016})}\BibitemShut {NoStop}%
\bibitem [{\citenamefont {Sugden}\ \emph {et~al.}(2022)\citenamefont {Sugden}, \citenamefont {Francia}, \citenamefont {Jensen}, \citenamefont {Adjiman},\ and\ \citenamefont {Salvalaglio}}]{sugden2022rationalising}%
  \BibitemOpen
  \bibfield  {author} {\bibinfo {author} {\bibfnamefont {I.~J.}\ \bibnamefont {Sugden}}, \bibinfo {author} {\bibfnamefont {N.~F.}\ \bibnamefont {Francia}}, \bibinfo {author} {\bibfnamefont {T.}~\bibnamefont {Jensen}}, \bibinfo {author} {\bibfnamefont {C.~S.}\ \bibnamefont {Adjiman}},\ and\ \bibinfo {author} {\bibfnamefont {M.}~\bibnamefont {Salvalaglio}},\ }\bibfield  {title} {\bibinfo {title} {Rationalising the difference in crystallisability of two sulflowers using efficient in silico methods},\ }\href {https://doi.org/https://doi.org/10.1039/D2CE00942K} {\bibfield  {journal} {\bibinfo  {journal} {CrystEngComm}\ }\textbf {\bibinfo {volume} {24}},\ \bibinfo {pages} {6830} (\bibinfo {year} {2022})}\BibitemShut {NoStop}%
\bibitem [{\citenamefont {Song}\ \emph {et~al.}(2020)\citenamefont {Song}, \citenamefont {Vogt-Maranto}, \citenamefont {Wiscons}, \citenamefont {Matzger},\ and\ \citenamefont {Tuckerman}}]{song2020generating}%
  \BibitemOpen
  \bibfield  {author} {\bibinfo {author} {\bibfnamefont {H.}~\bibnamefont {Song}}, \bibinfo {author} {\bibfnamefont {L.}~\bibnamefont {Vogt-Maranto}}, \bibinfo {author} {\bibfnamefont {R.}~\bibnamefont {Wiscons}}, \bibinfo {author} {\bibfnamefont {A.~J.}\ \bibnamefont {Matzger}},\ and\ \bibinfo {author} {\bibfnamefont {M.~E.}\ \bibnamefont {Tuckerman}},\ }\bibfield  {title} {\bibinfo {title} {Generating cocrystal polymorphs with information entropy driven by molecular dynamics-based enhanced sampling},\ }\href {https://doi.org/https://doi.org/10.1021/acs.jpclett.0c02647} {\bibfield  {journal} {\bibinfo  {journal} {J. Phys. Chem. Lett.}\ }\textbf {\bibinfo {volume} {11}},\ \bibinfo {pages} {9751} (\bibinfo {year} {2020})}\BibitemShut {NoStop}%
\bibitem [{\citenamefont {Francia}\ \emph {et~al.}(2021)\citenamefont {Francia}, \citenamefont {Price},\ and\ \citenamefont {Salvalaglio}}]{francia2021reducing}%
  \BibitemOpen
  \bibfield  {author} {\bibinfo {author} {\bibfnamefont {N.~F.}\ \bibnamefont {Francia}}, \bibinfo {author} {\bibfnamefont {L.~S.}\ \bibnamefont {Price}},\ and\ \bibinfo {author} {\bibfnamefont {M.}~\bibnamefont {Salvalaglio}},\ }\bibfield  {title} {\bibinfo {title} {Reducing crystal structure overprediction of ibuprofen with large scale molecular dynamics simulations},\ }\href {https://doi.org/https://doi.org/10.1039/D1CE00616A} {\bibfield  {journal} {\bibinfo  {journal} {CrystEngComm}\ }\textbf {\bibinfo {volume} {23}},\ \bibinfo {pages} {5575} (\bibinfo {year} {2021})}\BibitemShut {NoStop}%
\bibitem [{\citenamefont {Krummenacher}\ \emph {et~al.}(2024{\natexlab{a}})\citenamefont {Krummenacher}, \citenamefont {Tayfuroglu}, \citenamefont {Finkler}, \citenamefont {Huber},\ and\ \citenamefont {Goedecker}}]{krummenacher2024entropic}%
  \BibitemOpen
  \bibfield  {author} {\bibinfo {author} {\bibfnamefont {M.}~\bibnamefont {Krummenacher}}, \bibinfo {author} {\bibfnamefont {O.}~\bibnamefont {Tayfuroglu}}, \bibinfo {author} {\bibfnamefont {J.}~\bibnamefont {Finkler}}, \bibinfo {author} {\bibfnamefont {H.}~\bibnamefont {Huber}},\ and\ \bibinfo {author} {\bibfnamefont {S.}~\bibnamefont {Goedecker}},\ }\bibfield  {title} {\bibinfo {title} {Entropic stabilization of a structurally tolerant phase: The ionic phase of lithium alanate},\ }\bibfield  {journal} {\bibinfo  {journal} {Research Square Preprints}\ }\href {https://doi.org/https://doi.org/10.21203/rs.3.rs-4318358/v1} {https://doi.org/10.21203/rs.3.rs-4318358/v1} (\bibinfo {year} {2024}{\natexlab{a}})\BibitemShut {NoStop}%
\bibitem [{\citenamefont {Wales}(2004)}]{Walesbook}%
  \BibitemOpen
  \bibfield  {author} {\bibinfo {author} {\bibfnamefont {D.}~\bibnamefont {Wales}},\ }\href@noop {} {\emph {\bibinfo {title} {Energy Landscapes: Applications to Clusters, Biomolecules and Glasses}}},\ \bibinfo {edition} {1st}\ ed.\ (\bibinfo  {publisher} {Cambridge University Press},\ \bibinfo {year} {2004})\BibitemShut {NoStop}%
\bibitem [{\citenamefont {Bar}\ and\ \citenamefont {Bernstein}(1982)}]{bar1982molecular}%
  \BibitemOpen
  \bibfield  {author} {\bibinfo {author} {\bibfnamefont {I.}~\bibnamefont {Bar}}\ and\ \bibinfo {author} {\bibfnamefont {J.}~\bibnamefont {Bernstein}},\ }\bibfield  {title} {\bibinfo {title} {Molecular conformation and electronic structure. vi. the structure of p-methyl-n-(p-methylbenzylidene) aniline (form i)},\ }\href {https://doi.org/https://doi.org/10.1107/S0567740882002209} {\bibfield  {journal} {\bibinfo  {journal} {Acta Crystallographica Section B: Structural Crystallography and Crystal Chemistry}\ }\textbf {\bibinfo {volume} {38}},\ \bibinfo {pages} {121} (\bibinfo {year} {1982})}\BibitemShut {NoStop}%
\bibitem [{\citenamefont {Bar}\ and\ \citenamefont {Bernstein}(1977)}]{bar1977molecular}%
  \BibitemOpen
  \bibfield  {author} {\bibinfo {author} {\bibfnamefont {I.}~\bibnamefont {Bar}}\ and\ \bibinfo {author} {\bibfnamefont {J.}~\bibnamefont {Bernstein}},\ }\bibfield  {title} {\bibinfo {title} {Molecular conformation and electronic structure. v. the crystal and molecular structure of n-(p-methylbenzylidine)-p-methylaniline (form ii)},\ }\href {https://doi.org/https://doi.org/10.1107/S0567740877006980} {\bibfield  {journal} {\bibinfo  {journal} {Acta Crystallographica Section B: Structural Crystallography and Crystal Chemistry}\ }\textbf {\bibinfo {volume} {33}},\ \bibinfo {pages} {1738} (\bibinfo {year} {1977})}\BibitemShut {NoStop}%
\bibitem [{\citenamefont {Bernstein}\ \emph {et~al.}(1976)\citenamefont {Bernstein}, \citenamefont {Bar},\ and\ \citenamefont {Christensen}}]{bernstein1976molecular}%
  \BibitemOpen
  \bibfield  {author} {\bibinfo {author} {\bibfnamefont {J.}~\bibnamefont {Bernstein}}, \bibinfo {author} {\bibfnamefont {I.}~\bibnamefont {Bar}},\ and\ \bibinfo {author} {\bibfnamefont {A.}~\bibnamefont {Christensen}},\ }\bibfield  {title} {\bibinfo {title} {Molecular conformation and electronic structure. iv. p-(n-methylbenzylidene)-p-methylaniline (form iii)},\ }\href {https://doi.org/https://doi.org/10.1107/S0567740876006006} {\bibfield  {journal} {\bibinfo  {journal} {Acta Crystallographica Section B: Structural Crystallography and Crystal Chemistry}\ }\textbf {\bibinfo {volume} {32}},\ \bibinfo {pages} {1609} (\bibinfo {year} {1976})}\BibitemShut {NoStop}%
\bibitem [{\citenamefont {Goedecker}(2004)}]{goedecker:2004}%
  \BibitemOpen
  \bibfield  {author} {\bibinfo {author} {\bibfnamefont {S.}~\bibnamefont {Goedecker}},\ }\bibfield  {title} {\bibinfo {title} {Minima hopping: An efficient search method for the global minimum of the potential energy surface of complex molecular systems},\ }\href {https://doi.org/https://doi.org/10.1063/1.1724816} {\bibfield  {journal} {\bibinfo  {journal} {J. Chem. Phys.}\ }\textbf {\bibinfo {volume} {120}},\ \bibinfo {pages} {9911} (\bibinfo {year} {2004})}\BibitemShut {NoStop}%
\bibitem [{\citenamefont {Amsler}\ and\ \citenamefont {Goedecker}(2010)}]{amsler}%
  \BibitemOpen
  \bibfield  {author} {\bibinfo {author} {\bibfnamefont {M.}~\bibnamefont {Amsler}}\ and\ \bibinfo {author} {\bibfnamefont {S.}~\bibnamefont {Goedecker}},\ }\bibfield  {title} {\bibinfo {title} {Crystal structure prediction using the minima hopping method},\ }\href {https://doi.org/10.1063/1.3512900} {\bibfield  {journal} {\bibinfo  {journal} {The Journal of Chemical Physics}\ }\textbf {\bibinfo {volume} {133}},\ \bibinfo {pages} {224104} (\bibinfo {year} {2010})},\ \Eprint {https://arxiv.org/abs/https://pubs.aip.org/aip/jcp/article-pdf/doi/10.1063/1.3512900/15433393/224104\_1\_online.pdf} {https://pubs.aip.org/aip/jcp/article-pdf/doi/10.1063/1.3512900/15433393/224104\_1\_online.pdf} \BibitemShut {NoStop}%
\bibitem [{\citenamefont {Krummenacher}\ \emph {et~al.}(2024{\natexlab{b}})\citenamefont {Krummenacher}, \citenamefont {Gubler}, \citenamefont {Finkler}, \citenamefont {Huber}, \citenamefont {Sommer-J{\"o}rgensen},\ and\ \citenamefont {Goedecker}}]{krummenacher2024performing}%
  \BibitemOpen
  \bibfield  {author} {\bibinfo {author} {\bibfnamefont {M.}~\bibnamefont {Krummenacher}}, \bibinfo {author} {\bibfnamefont {M.}~\bibnamefont {Gubler}}, \bibinfo {author} {\bibfnamefont {J.~A.}\ \bibnamefont {Finkler}}, \bibinfo {author} {\bibfnamefont {H.}~\bibnamefont {Huber}}, \bibinfo {author} {\bibfnamefont {M.}~\bibnamefont {Sommer-J{\"o}rgensen}},\ and\ \bibinfo {author} {\bibfnamefont {S.}~\bibnamefont {Goedecker}},\ }\bibfield  {title} {\bibinfo {title} {Performing highly efficient minima hopping structure predictions using the atomic simulation environment (ase)},\ }\href {https://doi.org/https://doi.org/10.1016/j.softx.2024.101632} {\bibfield  {journal} {\bibinfo  {journal} {SoftwareX}\ }\textbf {\bibinfo {volume} {25}},\ \bibinfo {pages} {101632} (\bibinfo {year} {2024}{\natexlab{b}})}\BibitemShut {NoStop}%
\bibitem [{\citenamefont {Gubler}\ \emph {et~al.}(2023)\citenamefont {Gubler}, \citenamefont {Krummenacher}, \citenamefont {Huber},\ and\ \citenamefont {Goedecker}}]{gubler2023efficient}%
  \BibitemOpen
  \bibfield  {author} {\bibinfo {author} {\bibfnamefont {M.}~\bibnamefont {Gubler}}, \bibinfo {author} {\bibfnamefont {M.}~\bibnamefont {Krummenacher}}, \bibinfo {author} {\bibfnamefont {H.}~\bibnamefont {Huber}},\ and\ \bibinfo {author} {\bibfnamefont {S.}~\bibnamefont {Goedecker}},\ }\bibfield  {title} {\bibinfo {title} {Efficient variable cell shape geometry optimization},\ }\href {https://doi.org/https://doi.org/10.1016/j.jcpx.2023.100131} {\bibfield  {journal} {\bibinfo  {journal} {J. Comput. Phys. X}\ }\textbf {\bibinfo {volume} {17}},\ \bibinfo {pages} {100131} (\bibinfo {year} {2023})}\BibitemShut {NoStop}%
\bibitem [{\citenamefont {Ghasemi}\ \emph {et~al.}(2010)\citenamefont {Ghasemi}, \citenamefont {Amsler}, \citenamefont {Hennig}, \citenamefont {Roy}, \citenamefont {Goedecker}, \citenamefont {Lenosky}, \citenamefont {Umrigar}, \citenamefont {Genovese}, \citenamefont {Morishita},\ and\ \citenamefont {Nishio}}]{PhysRevB.81.214107}%
  \BibitemOpen
  \bibfield  {author} {\bibinfo {author} {\bibfnamefont {S.~A.}\ \bibnamefont {Ghasemi}}, \bibinfo {author} {\bibfnamefont {M.}~\bibnamefont {Amsler}}, \bibinfo {author} {\bibfnamefont {R.~G.}\ \bibnamefont {Hennig}}, \bibinfo {author} {\bibfnamefont {S.}~\bibnamefont {Roy}}, \bibinfo {author} {\bibfnamefont {S.}~\bibnamefont {Goedecker}}, \bibinfo {author} {\bibfnamefont {T.~J.}\ \bibnamefont {Lenosky}}, \bibinfo {author} {\bibfnamefont {C.~J.}\ \bibnamefont {Umrigar}}, \bibinfo {author} {\bibfnamefont {L.}~\bibnamefont {Genovese}}, \bibinfo {author} {\bibfnamefont {T.}~\bibnamefont {Morishita}},\ and\ \bibinfo {author} {\bibfnamefont {K.}~\bibnamefont {Nishio}},\ }\bibfield  {title} {\bibinfo {title} {Energy landscape of silicon systems and its description by force fields, tight binding schemes, density functional methods, and quantum monte carlo methods},\ }\href {https://doi.org/10.1103/PhysRevB.81.214107} {\bibfield  {journal} {\bibinfo  {journal} {Phys. Rev. B}\ }\textbf {\bibinfo {volume} {81}},\
  \bibinfo {pages} {214107} (\bibinfo {year} {2010})}\BibitemShut {NoStop}%
\bibitem [{\citenamefont {Kozhevnikov}\ \emph {et~al.}(2021)\citenamefont {Kozhevnikov}, \citenamefont {Taillefumier},\ and\ \citenamefont {Pintarelli}}]{sirius}%
  \BibitemOpen
  \bibfield  {author} {\bibinfo {author} {\bibfnamefont {A.}~\bibnamefont {Kozhevnikov}}, \bibinfo {author} {\bibfnamefont {M.}~\bibnamefont {Taillefumier}},\ and\ \bibinfo {author} {\bibfnamefont {S.}~\bibnamefont {Pintarelli}},\ }\href@noop {} {\bibinfo {title} {Sirius}},\ \bibinfo {howpublished} {\url{https://github.com/electronic-structure/SIRIUS}} (\bibinfo {year} {2021})\BibitemShut {NoStop}%
\bibitem [{\citenamefont {Goedecker}\ \emph {et~al.}(1996)\citenamefont {Goedecker}, \citenamefont {Teter},\ and\ \citenamefont {Hutter}}]{goedecker1996separable}%
  \BibitemOpen
  \bibfield  {author} {\bibinfo {author} {\bibfnamefont {S.}~\bibnamefont {Goedecker}}, \bibinfo {author} {\bibfnamefont {M.}~\bibnamefont {Teter}},\ and\ \bibinfo {author} {\bibfnamefont {J.}~\bibnamefont {Hutter}},\ }\bibfield  {title} {\bibinfo {title} {Separable dual-space gaussian pseudopotentials},\ }\href {https://doi.org/https://doi.org/10.1103/PhysRevB.54.1703} {\bibfield  {journal} {\bibinfo  {journal} {Phys. Rev. B}\ }\textbf {\bibinfo {volume} {54}},\ \bibinfo {pages} {1703} (\bibinfo {year} {1996})}\BibitemShut {NoStop}%
\bibitem [{\citenamefont {Price}\ \emph {et~al.}(2021)\citenamefont {Price}, \citenamefont {Bryenton},\ and\ \citenamefont {Johnson}}]{price2021requirements}%
  \BibitemOpen
  \bibfield  {author} {\bibinfo {author} {\bibfnamefont {A.~J.}\ \bibnamefont {Price}}, \bibinfo {author} {\bibfnamefont {K.~R.}\ \bibnamefont {Bryenton}},\ and\ \bibinfo {author} {\bibfnamefont {E.~R.}\ \bibnamefont {Johnson}},\ }\bibfield  {title} {\bibinfo {title} {Requirements for an accurate dispersion-corrected density functional},\ }\bibfield  {journal} {\bibinfo  {journal} {J. Chem. Phys.}\ }\textbf {\bibinfo {volume} {154}},\ \href {https://doi.org/https://doi.org/10.1063/5.0050993} {https://doi.org/10.1063/5.0050993} (\bibinfo {year} {2021})\BibitemShut {NoStop}%
\bibitem [{\citenamefont {Beran}(2016)}]{beran2016modeling}%
  \BibitemOpen
  \bibfield  {author} {\bibinfo {author} {\bibfnamefont {G.~J.}\ \bibnamefont {Beran}},\ }\bibfield  {title} {\bibinfo {title} {Modeling polymorphic molecular crystals with electronic structure theory},\ }\href {https://doi.org/https://doi.org/10.1021/acs.chemrev.5b00648} {\bibfield  {journal} {\bibinfo  {journal} {Chem. Rev.}\ }\textbf {\bibinfo {volume} {116}},\ \bibinfo {pages} {5567} (\bibinfo {year} {2016})}\BibitemShut {NoStop}%
\bibitem [{\citenamefont {Caldeweyher}\ \emph {et~al.}(2019)\citenamefont {Caldeweyher}, \citenamefont {Ehlert}, \citenamefont {Hansen}, \citenamefont {Neugebauer}, \citenamefont {Spicher}, \citenamefont {Bannwarth},\ and\ \citenamefont {Grimme}}]{caldeweyher2019generally}%
  \BibitemOpen
  \bibfield  {author} {\bibinfo {author} {\bibfnamefont {E.}~\bibnamefont {Caldeweyher}}, \bibinfo {author} {\bibfnamefont {S.}~\bibnamefont {Ehlert}}, \bibinfo {author} {\bibfnamefont {A.}~\bibnamefont {Hansen}}, \bibinfo {author} {\bibfnamefont {H.}~\bibnamefont {Neugebauer}}, \bibinfo {author} {\bibfnamefont {S.}~\bibnamefont {Spicher}}, \bibinfo {author} {\bibfnamefont {C.}~\bibnamefont {Bannwarth}},\ and\ \bibinfo {author} {\bibfnamefont {S.}~\bibnamefont {Grimme}},\ }\bibfield  {title} {\bibinfo {title} {A generally applicable atomic-charge dependent london dispersion correction},\ }\bibfield  {journal} {\bibinfo  {journal} {J. Chem. Phys.}\ }\textbf {\bibinfo {volume} {150}},\ \href {https://doi.org/https://doi.org/10.1063/1.5090222} {https://doi.org/10.1063/1.5090222} (\bibinfo {year} {2019})\BibitemShut {NoStop}%
\bibitem [{\citenamefont {Caldeweyher}\ \emph {et~al.}(2020)\citenamefont {Caldeweyher}, \citenamefont {Mewes}, \citenamefont {Ehlert},\ and\ \citenamefont {Grimme}}]{caldeweyher2020extension}%
  \BibitemOpen
  \bibfield  {author} {\bibinfo {author} {\bibfnamefont {E.}~\bibnamefont {Caldeweyher}}, \bibinfo {author} {\bibfnamefont {J.-M.}\ \bibnamefont {Mewes}}, \bibinfo {author} {\bibfnamefont {S.}~\bibnamefont {Ehlert}},\ and\ \bibinfo {author} {\bibfnamefont {S.}~\bibnamefont {Grimme}},\ }\bibfield  {title} {\bibinfo {title} {Extension and evaluation of the d4 london-dispersion model for periodic systems},\ }\href {https://doi.org/https://doi.org/10.1039/D0CP00502A} {\bibfield  {journal} {\bibinfo  {journal} {PCCP}\ }\textbf {\bibinfo {volume} {22}},\ \bibinfo {pages} {8499} (\bibinfo {year} {2020})}\BibitemShut {NoStop}%
\bibitem [{\citenamefont {Perdew}\ \emph {et~al.}(1996)\citenamefont {Perdew}, \citenamefont {Burke},\ and\ \citenamefont {Ernzerhof}}]{perdew1996generalized}%
  \BibitemOpen
  \bibfield  {author} {\bibinfo {author} {\bibfnamefont {J.~P.}\ \bibnamefont {Perdew}}, \bibinfo {author} {\bibfnamefont {K.}~\bibnamefont {Burke}},\ and\ \bibinfo {author} {\bibfnamefont {M.}~\bibnamefont {Ernzerhof}},\ }\bibfield  {title} {\bibinfo {title} {Generalized gradient approximation made simple},\ }\href {https://doi.org/https://doi.org/10.1103/PhysRevLett.77.3865} {\bibfield  {journal} {\bibinfo  {journal} {Phys. Rev. Lett.}\ }\textbf {\bibinfo {volume} {77}},\ \bibinfo {pages} {3865} (\bibinfo {year} {1996})}\BibitemShut {NoStop}%
\bibitem [{\citenamefont {Batzner}\ \emph {et~al.}(2022)\citenamefont {Batzner}, \citenamefont {Musaelian}, \citenamefont {Sun}, \citenamefont {Geiger}, \citenamefont {Mailoa}, \citenamefont {Kornbluth}, \citenamefont {Molinari}, \citenamefont {Smidt},\ and\ \citenamefont {Kozinsky}}]{batzner20223}%
  \BibitemOpen
  \bibfield  {author} {\bibinfo {author} {\bibfnamefont {S.}~\bibnamefont {Batzner}}, \bibinfo {author} {\bibfnamefont {A.}~\bibnamefont {Musaelian}}, \bibinfo {author} {\bibfnamefont {L.}~\bibnamefont {Sun}}, \bibinfo {author} {\bibfnamefont {M.}~\bibnamefont {Geiger}}, \bibinfo {author} {\bibfnamefont {J.~P.}\ \bibnamefont {Mailoa}}, \bibinfo {author} {\bibfnamefont {M.}~\bibnamefont {Kornbluth}}, \bibinfo {author} {\bibfnamefont {N.}~\bibnamefont {Molinari}}, \bibinfo {author} {\bibfnamefont {T.~E.}\ \bibnamefont {Smidt}},\ and\ \bibinfo {author} {\bibfnamefont {B.}~\bibnamefont {Kozinsky}},\ }\bibfield  {title} {\bibinfo {title} {E (3)-equivariant graph neural networks for data-efficient and accurate interatomic potentials},\ }\href {https://doi.org/https://doi.org/10.1038/s41467-022-29939-5} {\bibfield  {journal} {\bibinfo  {journal} {Nature communications}\ }\textbf {\bibinfo {volume} {13}},\ \bibinfo {pages} {2453} (\bibinfo {year} {2022})}\BibitemShut {NoStop}%
\bibitem [{\citenamefont {Larsen}\ \emph {et~al.}(2017)\citenamefont {Larsen}, \citenamefont {Mortensen}, \citenamefont {Blomqvist}, \citenamefont {Castelli}, \citenamefont {Christensen}, \citenamefont {Du{\l}ak}, \citenamefont {Friis}, \citenamefont {Groves}, \citenamefont {Hammer}, \citenamefont {Hargus} \emph {et~al.}}]{larsen2017atomic}%
  \BibitemOpen
  \bibfield  {author} {\bibinfo {author} {\bibfnamefont {A.~H.}\ \bibnamefont {Larsen}}, \bibinfo {author} {\bibfnamefont {J.~J.}\ \bibnamefont {Mortensen}}, \bibinfo {author} {\bibfnamefont {J.}~\bibnamefont {Blomqvist}}, \bibinfo {author} {\bibfnamefont {I.~E.}\ \bibnamefont {Castelli}}, \bibinfo {author} {\bibfnamefont {R.}~\bibnamefont {Christensen}}, \bibinfo {author} {\bibfnamefont {M.}~\bibnamefont {Du{\l}ak}}, \bibinfo {author} {\bibfnamefont {J.}~\bibnamefont {Friis}}, \bibinfo {author} {\bibfnamefont {M.~N.}\ \bibnamefont {Groves}}, \bibinfo {author} {\bibfnamefont {B.}~\bibnamefont {Hammer}}, \bibinfo {author} {\bibfnamefont {C.}~\bibnamefont {Hargus}}, \emph {et~al.},\ }\bibfield  {title} {\bibinfo {title} {The atomic simulation environment—a python library for working with atoms},\ }\href {https://doi.org/10.1088/1361-648X/aa680e} {\bibfield  {journal} {\bibinfo  {journal} {J. Condens. Matter Phys.}\ }\textbf {\bibinfo {volume} {29}},\ \bibinfo {pages} {273002} (\bibinfo {year}
  {2017})}\BibitemShut {NoStop}%
\bibitem [{\citenamefont {Price}(2018)}]{price2018control}%
  \BibitemOpen
  \bibfield  {author} {\bibinfo {author} {\bibfnamefont {S.~L.}\ \bibnamefont {Price}},\ }\bibfield  {title} {\bibinfo {title} {Control and prediction of the organic solid state: a challenge to theory and experiment},\ }\href {https://doi.org/https://doi.org/10.1098/rspa.2018.0351} {\bibfield  {journal} {\bibinfo  {journal} {Proceedings of the Royal Society A: Mathematical, Physical and Engineering Sciences}\ }\textbf {\bibinfo {volume} {474}},\ \bibinfo {pages} {20180351} (\bibinfo {year} {2018})}\BibitemShut {NoStop}%
\bibitem [{\citenamefont {Price}\ and\ \citenamefont {Reutzel-Edens}(2016)}]{price2016potential}%
  \BibitemOpen
  \bibfield  {author} {\bibinfo {author} {\bibfnamefont {S.~L.}\ \bibnamefont {Price}}\ and\ \bibinfo {author} {\bibfnamefont {S.~M.}\ \bibnamefont {Reutzel-Edens}},\ }\bibfield  {title} {\bibinfo {title} {The potential of computed crystal energy landscapes to aid solid-form development},\ }\href {https://doi.org/https://doi.org/10.1016/j.drudis.2016.01.014} {\bibfield  {journal} {\bibinfo  {journal} {Drug Discovery Today}\ }\textbf {\bibinfo {volume} {21}},\ \bibinfo {pages} {912} (\bibinfo {year} {2016})}\BibitemShut {NoStop}%
\bibitem [{\citenamefont {Schaefer}\ \emph {et~al.}(2014)\citenamefont {Schaefer}, \citenamefont {Mohr}, \citenamefont {Amsler},\ and\ \citenamefont {Goedecker}}]{schaefer2014minima}%
  \BibitemOpen
  \bibfield  {author} {\bibinfo {author} {\bibfnamefont {B.}~\bibnamefont {Schaefer}}, \bibinfo {author} {\bibfnamefont {S.}~\bibnamefont {Mohr}}, \bibinfo {author} {\bibfnamefont {M.}~\bibnamefont {Amsler}},\ and\ \bibinfo {author} {\bibfnamefont {S.}~\bibnamefont {Goedecker}},\ }\bibfield  {title} {\bibinfo {title} {Minima hopping guided path search: An efficient method for finding complex chemical reaction pathways},\ }\bibfield  {journal} {\bibinfo  {journal} {J. Chem. Phys.}\ }\textbf {\bibinfo {volume} {140}},\ \href {https://doi.org/https://doi.org/10.1063/1.4878944} {https://doi.org/10.1063/1.4878944} (\bibinfo {year} {2014})\BibitemShut {NoStop}%
\bibitem [{\citenamefont {Sommer-J{\"o}rgensen}\ and\ \citenamefont {Goedecker}(2024)}]{sommer2024compass}%
  \BibitemOpen
  \bibfield  {author} {\bibinfo {author} {\bibfnamefont {M.}~\bibnamefont {Sommer-J{\"o}rgensen}}\ and\ \bibinfo {author} {\bibfnamefont {S.}~\bibnamefont {Goedecker}},\ }\bibfield  {title} {\bibinfo {title} {Compass: Double-ended saddle point search as a constrained optimization problem},\ }\bibfield  {journal} {\bibinfo  {journal} {J. Chem. Phys.}\ }\textbf {\bibinfo {volume} {160}},\ \href {https://doi.org/https://doi.org/10.1063/5.0186903} {https://doi.org/10.1063/5.0186903} (\bibinfo {year} {2024})\BibitemShut {NoStop}%
\bibitem [{\citenamefont {Zhu}\ \emph {et~al.}(2016)\citenamefont {Zhu}, \citenamefont {Amsler}, \citenamefont {Fuhrer}, \citenamefont {Schaefer}, \citenamefont {Faraji}, \citenamefont {Rostami}, \citenamefont {Ghasemi}, \citenamefont {Sadeghi}, \citenamefont {Grauzinyte}, \citenamefont {Wolverton} \emph {et~al.}}]{zhu2016fingerprint}%
  \BibitemOpen
  \bibfield  {author} {\bibinfo {author} {\bibfnamefont {L.}~\bibnamefont {Zhu}}, \bibinfo {author} {\bibfnamefont {M.}~\bibnamefont {Amsler}}, \bibinfo {author} {\bibfnamefont {T.}~\bibnamefont {Fuhrer}}, \bibinfo {author} {\bibfnamefont {B.}~\bibnamefont {Schaefer}}, \bibinfo {author} {\bibfnamefont {S.}~\bibnamefont {Faraji}}, \bibinfo {author} {\bibfnamefont {S.}~\bibnamefont {Rostami}}, \bibinfo {author} {\bibfnamefont {S.~A.}\ \bibnamefont {Ghasemi}}, \bibinfo {author} {\bibfnamefont {A.}~\bibnamefont {Sadeghi}}, \bibinfo {author} {\bibfnamefont {M.}~\bibnamefont {Grauzinyte}}, \bibinfo {author} {\bibfnamefont {C.}~\bibnamefont {Wolverton}}, \emph {et~al.},\ }\bibfield  {title} {\bibinfo {title} {A fingerprint based metric for measuring similarities of crystalline structures},\ }\href {https://doi.org/https://doi.org/10.1063/1.4940026} {\bibfield  {journal} {\bibinfo  {journal} {J. Chem. Phys.}\ }\textbf {\bibinfo {volume} {144}},\ \bibinfo {pages} {034203} (\bibinfo {year} {2016})}\BibitemShut
  {NoStop}%
\bibitem [{\citenamefont {Thompson}\ \emph {et~al.}(2022)\citenamefont {Thompson}, \citenamefont {Aktulga}, \citenamefont {Berger}, \citenamefont {Bolintineanu}, \citenamefont {Brown}, \citenamefont {Crozier}, \citenamefont {{in 't Veld}}, \citenamefont {Kohlmeyer}, \citenamefont {Moore}, \citenamefont {Nguyen}, \citenamefont {Shan}, \citenamefont {Stevens}, \citenamefont {Tranchida}, \citenamefont {Trott},\ and\ \citenamefont {Plimpton}}]{thompson2022lammps}%
  \BibitemOpen
  \bibfield  {author} {\bibinfo {author} {\bibfnamefont {A.~P.}\ \bibnamefont {Thompson}}, \bibinfo {author} {\bibfnamefont {H.~M.}\ \bibnamefont {Aktulga}}, \bibinfo {author} {\bibfnamefont {R.}~\bibnamefont {Berger}}, \bibinfo {author} {\bibfnamefont {D.~S.}\ \bibnamefont {Bolintineanu}}, \bibinfo {author} {\bibfnamefont {W.~M.}\ \bibnamefont {Brown}}, \bibinfo {author} {\bibfnamefont {P.~S.}\ \bibnamefont {Crozier}}, \bibinfo {author} {\bibfnamefont {P.~J.}\ \bibnamefont {{in 't Veld}}}, \bibinfo {author} {\bibfnamefont {A.}~\bibnamefont {Kohlmeyer}}, \bibinfo {author} {\bibfnamefont {S.~G.}\ \bibnamefont {Moore}}, \bibinfo {author} {\bibfnamefont {T.~D.}\ \bibnamefont {Nguyen}}, \bibinfo {author} {\bibfnamefont {R.}~\bibnamefont {Shan}}, \bibinfo {author} {\bibfnamefont {M.~J.}\ \bibnamefont {Stevens}}, \bibinfo {author} {\bibfnamefont {J.}~\bibnamefont {Tranchida}}, \bibinfo {author} {\bibfnamefont {C.}~\bibnamefont {Trott}},\ and\ \bibinfo {author} {\bibfnamefont {S.~J.}\ \bibnamefont {Plimpton}},\
  }\bibfield  {title} {\bibinfo {title} {Lammps - a flexible simulation tool for particle-based materials modeling at the atomic, meso, and continuum scales},\ }\href {https://doi.org/https://doi.org/10.1016/j.cpc.2021.108171} {\bibfield  {journal} {\bibinfo  {journal} {Comput. Phys. Commun.}\ }\textbf {\bibinfo {volume} {271}},\ \bibinfo {pages} {108171} (\bibinfo {year} {2022})}\BibitemShut {NoStop}%
\end{thebibliography}%


\end{document}